\documentclass[sigconf,authorversion]{acmart} %

\AtBeginDocument{%
  \providecommand\BibTeX{{%
    \normalfont B\kern-0.5em{\scshape i\kern-0.25em b}\kern-0.8em\TeX}}}

\usepackage[capitalise]{cleveref}
\crefname{section}{Sec.}{Secs.} %
\Crefname{section}{Sec.}{Secs.} %

\usepackage{xspace}
\usepackage{soul}

\usepackage[inline]{enumitem}
\usepackage{gensymb}
\usepackage{pifont}
\usepackage{fancybox}

\usepackage{ifthen}
\usepackage{balance}

\newcommand{\SystemName}{\textsc{CoCreatAR}\xspace}
\newcommand{\insitu}[1][-]{{in\ifthenelse{\equal{#1}{-}}{-}{ }situ}\xspace}
\newcommand{\exsitu}[1][-]{{ex\ifthenelse{\equal{#1}{-}}{-}{ }situ}\xspace}
\newcommand{\Insitu}[1][-]{{In\ifthenelse{\equal{#1}{-}}{-}{ }situ}\xspace}
\newcommand{\Exsitu}[1][-]{{Ex\ifthenelse{\equal{#1}{-}}{-}{ }situ}\xspace}
\newcommand{\sync}{\textsc{Sync}\xspace}
\newcommand{\async}{\textsc{Async}\xspace}
\newcommand{\locA}{\textit{Location} \textsc{A}\xspace}
\newcommand{\locB}{\textit{Location} \textsc{B}\xspace}
\newcommand{\locMesh}{\textit{location mesh}\xspace} %
\newcommand{\locMeshes}{\textit{location meshes}\xspace} %

\newtheorem{hypothesis}{Hypothesis}

\newcommand*{\eg}{e.g.\@\xspace}

\copyrightyear{2025}
\acmYear{2025}
\setcopyright{rightsretained}
\acmConference[CHI '25]{CHI Conference on Human Factors in Computing Systems}{April 26-May 1, 2025}{Yokohama, Japan}
\acmBooktitle{CHI Conference on Human Factors in Computing Systems (CHI '25), April 26-May 1, 2025, Yokohama, Japan}\acmDOI{10.1145/3706598.3714274}
\acmISBN{979-8-4007-1394-1/25/04}

\begin{document}

\title[CoCreatAR: Enhancing Authoring of Outdoor AR Experiences Through Asymmetric Collaboration]{CoCreatAR: Enhancing Authoring of Outdoor Augmented Reality Experiences Through Asymmetric Collaboration}

\author{Nels Numan}
\affiliation{%
  \institution{Niantic Inc.}
  \country{}
  }
\affiliation{%
  \institution{University College London}
  \city{London}
  \country{United Kingdom}
  }
\email{nels.numan@ucl.ac.uk}

\author{Gabriel Brostow}
\affiliation{%
  \institution{Niantic Inc.}
  \country{}
  }
\affiliation{%
  \institution{University College London}
  \city{London}
  \country{United Kingdom}
  }
\email{g.brostow@cs.ucl.ac.uk}

\author{Suhyun Park}
\affiliation{%
  \institution{University College London}
  \city{London}
  \country{United Kingdom}
  }
\email{pppshyun1226@gmail.com}

\author{Simon Julier}
\affiliation{%
  \institution{University College London}
  \city{London}
  \country{United Kingdom}
  }
\email{s.julier@ucl.ac.uk}

\author{Anthony Steed}
\affiliation{%
  \institution{University College London}
  \city{London}
  \country{United Kingdom}
  }
\email{a.steed@ucl.ac.uk}

\author{Jessica Van Brummelen}
\affiliation{%
  \institution{Niantic Inc.}
  \city{London}
  \country{United Kingdom}
  }
\email{jvanbrummelen@nianticlabs.com}

\renewcommand{\shortauthors}{N. Numan, G. Brostow, S. Park, S. Julier, A. Steed, and J. Van Brummelen}

\begin{abstract}
Authoring site-specific outdoor augmented reality (AR) experiences requires a nuanced understanding of real-world context to create immersive and relevant content. Existing ex-situ authoring tools typically rely on static 3D models to represent spatial information. However, in our formative study ($n$=25), we identified key limitations of this approach: models are often outdated, incomplete, or insufficient for capturing critical factors such as safety considerations, user flow, and dynamic environmental changes. These issues necessitate frequent on-site visits and additional iterations, making the authoring process more time-consuming and resource-intensive. To mitigate these challenges, we introduce \SystemName, an asymmetric collaborative mixed reality authoring system that integrates the flexibility of ex-situ workflows with the immediate contextual awareness of in-situ authoring. We conducted an exploratory study ($n$=32) comparing \SystemName to an asynchronous workflow baseline, finding that it enhances engagement, creativity, and confidence in the authored output while also providing preliminary insights into its impact on task load. We conclude by discussing the implications of our findings for integrating real-world context into site-specific AR authoring systems.
\end{abstract}

\ccsdesc[500]{Human-centered computing~Mixed / augmented reality}
\ccsdesc[500]{Computing methodologies~Mixed / augmented reality}
\ccsdesc[300]{Software and its engineering~Virtual worlds software}

\keywords{collaborative mixed reality, augmented reality, co-creation, site-specific, authoring tools, reconstruction, context-aware systems}

\begin{teaserfigure}
  \includegraphics[width=.998\textwidth]{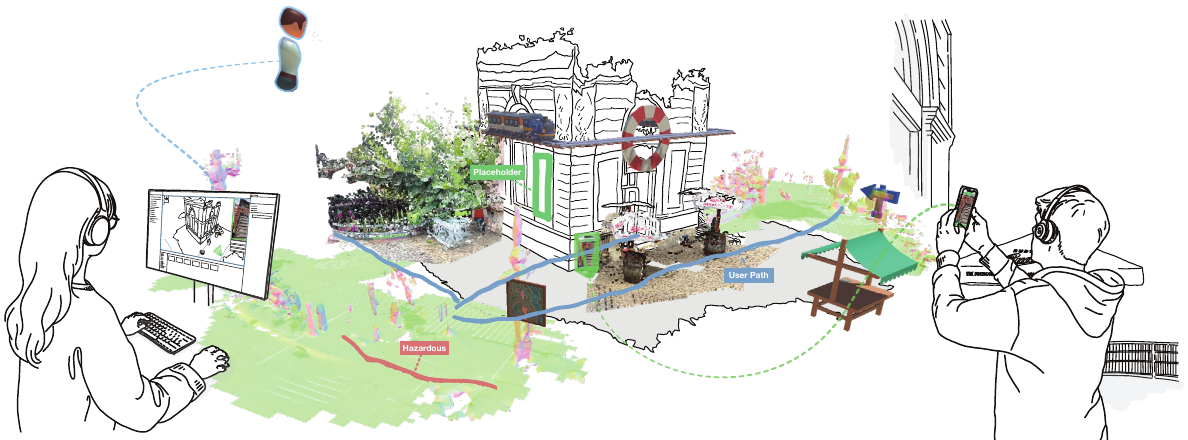}
  \caption{\SystemName enables \exsitu developers (left) to collaborate with \insitu users (right) in the development of site-specific AR experiences. While iterating on the design and refining the user experience, the \insitu user can share critical real-world contextual information with their remote partner synchronously. This includes audio and video, meshes, RGB-D point clouds, annotations, as well as feedback on usability and performance. This additional anchored context enables the \exsitu developer to refine the prototype effectively using familiar tools, without requiring repeated visits to the target site.}
  \label{fig:teaser}
\end{teaserfigure}

\maketitle

\section{Introduction}

Outdoor augmented reality (AR) experiences are often designed for specific locations, spanning a range of settings from controlled environments such as theme parks~\cite{LegolandAR2024} to dynamic public urban spaces~\cite{broadcastnow2023}. Developing these experiences is inherently complex because the user experience is closely intertwined with the physical environment. However, developers often work remotely, with only intermittent access to the target location.

\Exsitu (i.e., \emph{off-site}, \emph{remote}) development workflows typically rely on pre-captured 3D models of the environment obtained from prior scans or geospatial data to design and position AR content. Tools such as \textit{Google Geospatial Creator}~\cite{googleGeospatialCreator} and \textit{Niantic Remote Content Authoring}~\cite{lightship-ardk-niantic} facilitate this process by providing site-specific environmental representations at a large scale. Additionally, they offer a visual positioning system (VPS) to precisely align virtual content with the real-world target environment. While these tools effectively support development, their reliance on static representations often leads to missing critical contextual information.

\textit{Context}, broadly defined by \citet{abowdBetterUnderstandingContext1999} as any information characterizing the situation of relevant entities, is particularly important when creating experiences for outdoor settings. For instance, dynamic elements such as lighting conditions, moving objects (e.g., vehicles, temporary structures), and pedestrian activity are typically absent from pre-captured 3D models. This absence can lead to inconsistencies between virtual elements and the real environment, ultimately degrading the user experience~\cite{nebelingXRToolsWhere2022}. Consequently, developers frequently resort to repeated on-site visits, which can be costly, time-consuming, and logistically challenging.

\Insitu (i.e., \emph{on-site}) authoring tools, in contrast, enable AR content creation and testing directly within the target environment, providing immediate contextual awareness~\cite{langlotzSketchingWorldSitu2012,adobeAero,unityMars}. However, these tools often lack the flexibility and expressive power of remote development environments, particularly for large-scale outdoor experiences that require precise object placement, complex interactions, or considerations of user flow and safety. Moreover, mobile devices used for \insitu authoring often have limited computational resources and interaction capabilities, making it difficult to manage complex assets or scripting tasks~\cite{leeImmersiveAuthoringOfTangible2009,vargasgonzalezComparisonDesktopAugmented2019,loOffsiteOnsiteFlexible2021}.

To better understand the challenges of \exsitu authoring of site-specific outdoor AR experiences, we conducted a formative study comprising a survey ($n$=25) and follow-up interviews ($n$=5) with industry professionals. Our findings highlighted common issues, including missing or outdated environmental representations, time-consuming iterative testing cycles that necessitate site visits, and challenges in capturing and communicating contextual information among team members.

Building on these insights, we explored collaborative approaches that integrate the advantages of both \exsitu and \insitu authoring. By enabling synchronous collaboration between \exsitu developers and \insitu collaborators, we sought to address a set of major challenges we identified. Inspired by the concept of pair programming~\cite{hannayEffectivenessPairProgramming2009}, we structured collaborator roles so that the \insitu user captures and transmits real-time environmental context while the \exsitu developer remotely utilizes advanced authoring tools.

To support this workflow, we developed \SystemName, an asymmetric collaborative AR authoring system that facilitates real-time editing of site-specific AR content. \SystemName enables \exsitu developers to work synchronously with \insitu users by providing enhanced spatial information, anchored annotations, and communication tools to bridge the gap between remote development and the physical target environment.

We conducted a user study ($n$=32, in pairs) to compare the effect of synchronous (\SystemName) and asynchronous collaborative authoring approaches on task load, engagement, and confidence in the authored result. In the synchronous condition, \exsitu and \insitu collaborators used \SystemName to work together in real-time, whereas in the asynchronous condition, development proceeded sequentially without immediate interaction, reflecting current practices identified in our formative study. Our findings suggest that authoring with \SystemName improves the integration of real-world context, enhances developers' confidence in the accuracy and feasibility of their AR designs, and leads to greater engagement and creativity among team members. Overall, our contributions include:

\begin{itemize}
\item A formative study analyzing current developer workflows and the role of real-world context in site-specific outdoor AR experience development, based on a survey ($n$=25) and interviews ($n$=5) with industry professionals.
\item The design and implementation of \SystemName, an asymmetric collaborative system that supports real-time, site-specific outdoor AR content authoring by integrating \exsitu and \insitu roles.
\item An empirical evaluation comparing synchronous authoring with \SystemName and an asynchronous workflow baseline, demonstrating the benefits and trade-offs of each method through an exploratory user study ($n$=32).
\end{itemize}

\section{Related Work}
This section discusses related work on AR authoring tools, focusing on \insitu and \exsitu methods, collaborative workflows, and techniques for integrating dynamic environmental context.

\subsection{Leveraging Real-World Context in AR Experience Authoring}
AR authoring tools span a wide spectrum of fidelity levels, supporting different development stages and expertise levels. As categorized by \citet{nebelingTroubleAugmentedReality2018}, these range from low-fidelity prototyping tools that require little to no technical expertise~\cite{adobeAero,nebelingProtoARRapidPhysicalDigital2018,rajaramPaperTrailImmersive2022} to high-fidelity authoring tools, such as Unity~\cite{unity} and Unreal Engine~\cite{unreal}, which are commonly used for developing production-ready applications.

Despite their utility, existing AR authoring tools have challenges that limit their effectiveness. Several surveys and studies have identified key issues, including difficulties in spatial navigation, adapting to changing environmental contexts, and integrating context-specific considerations into AR experiences~\cite{kraussCurrentPracticesChallenges2021, ashtariCreatingAugmentedVirtual2020, kraussElementsXRPrototyping2022, nebelingXRToolsWhere2022}. A central limitation, as noted by \citet{nebelingXRToolsWhere2022}, is the insufficient incorporation of real-world context during authoring, which can disrupt user experience and immersion. To incorporate this context, authoring tools generally adopt one of two approaches: \textit{\insitu} authoring, which supports real-time creation in the target environment, or \textit{\exsitu} authoring, which relies on proxy representations or recordings of the target environment.

\subsubsection{\Insitu authoring}
Tools that incorporate \insitu interfaces, such as the mobile applications of Adobe Aero~\cite{adobeAero}, Reality Composer~\cite{realityComposer}, and Unity Mars~\cite{unityMars} and other prior work~\cite{zollmannComprehensibleInteractiveVisualizations2012,langlotzSketchingWorldSitu2012}, build on the \textit{WYXIWYG (What You Experience Is What You Get)} paradigm introduced by \citet{leeImmersiveAuthoringWhat2005}, which emphasizes the benefits of allowing users to experience and verify authored content \insitu[ ], such as the ability for immediate testing and adjustment. 

In a user study, \citet{leeImmersiveAuthoringOfTangible2009} found that their \insitu system allowed for efficient and precise arrangement of virtual content in real-world settings. However, they noted a key limitation: abstract tasks, such as programming behaviors, were better supported by traditional \exsitu desktop environments. \citet{langlotzSketchingWorldSitu2012} developed a mobile AR authoring system with a focus on enabling spontaneous content creation in unprepared environments, including outdoor settings. While highlighting the low barriers to entry \insitu authoring provides, they also emphasized the importance of integrating it with \exsitu desktop-based interfaces for refining the content. \citet{vargasgonzalezComparisonDesktopAugmented2019} compared \insitu AR and \exsitu desktop authoring tools in scenario-based training contexts and similarly found that while both approaches offered comparable usability and task completion times, desktop tools were perceived as more efficient, particularly for tasks requiring global context.

\subsubsection{\Exsitu authoring}
\Exsitu authoring tools, such as \textit{DART}~\cite{macintyreDARTToolkitRapid2004}, \textit{ScalAR}~\cite{qianScalARAuthoringSemantically2022}, \textit{DistanciAR}~\cite{wangDistanciARAuthoringSiteSpecific2021}, and \textit{Corsican Twin}~\cite{prouzeauCorsicanTwinAuthoring2020}, offer broader authoring capabilities than \insitu tools by providing 3D scene editors, scripting environments, flexible testing workflows, and asset integration. These tools represent environmental context using various forms of pre-captured data, including 3D models~\cite{prouzeauCorsicanTwinAuthoring2020, qianScalARAuthoringSemantically2022, cavalloCAVEARVRAuthoring2019, lightship-ardk-niantic, googleGeospatialCreator}, video~\cite{macintyreDARTToolkitRapid2004, leivaRapidoPrototypingInteractive2021}, sensor data~\cite{macintyreDARTToolkitRapid2004}, or $360\degree$ footage~\cite{nebelingProtoARRapidPhysicalDigital2018}. For example, \textit{DART}~\cite{macintyreDARTToolkitRapid2004} facilitates recording and playback for authoring by synchronizing real-time video and sensor data capture, while \textit{DistanciAR}~\cite{wangDistanciARAuthoringSiteSpecific2021} supports \insitu 3D reconstruction to capture environmental context and enables \exsitu authoring and testing through different model visualization modes. However, a key limitation of these approaches is their reliance on static environmental representations. These models capture only a fixed moment in time, falling short in accounting for dynamic changes. As a result, authoring errors and spatial misalignment may occur.

Recent \exsitu authoring tools for outdoor AR experiences, such as \textit{Niantic Remote Content Authoring}~\cite{lightship-ardk-niantic} and \textit{Google Geospatial Creator}~\cite{googleGeospatialCreator}, adopt similar approaches by providing 3D models of large outdoor sites. However, outdoor environments present even greater challenges due to a higher degree of dynamic change. Unlike indoor spaces, outdoor settings are continuously affected by shifting lighting, weather, seasons, and moving elements like people and vehicles. These dynamic factors significantly impact the accuracy and relevance of pre-captured data, making static environmental representations particularly problematic for authoring outdoor AR content. While \insitu testing can help mitigate some challenges, it is often constrained by logistical factors such as time, cost, and accessibility. Moreover, \insitu authoring of outdoor AR experiences is complicated by the scale and complexity of outdoor spaces~\cite{imottesjoIterativePrototypingUrban2020, numanOutdoorCollaborativeMixed2023, caoMobileAugmentedReality2023}, making \insitu adjustments more difficult to execute.

Overall, our analysis indicates that current AR authoring tools remain limited in bridging the gap between \exsitu flexibility and \insitu contextual awareness. \Exsitu tools rely on static proxies that fail to capture the dynamic nature of outdoor spaces, while \insitu tools, despite offering real-time context, lack the expressiveness and flexibility required for comprehensive authoring, as we found in our formative study (\cref{sec:formative-study}). This disconnect forces authors to switch between \exsitu and \insitu development, leading to inefficiencies and challenges in maintaining contextual accuracy. To address this gap, we propose \SystemName, a system that leverages the flexibility of \exsitu authoring while incorporating real-time environmental awareness from \insitu users. By enabling a synchronous collaborative workflow between \exsitu and \insitu roles, our approach supports contextually aware authoring without compromising the richness and advanced capabilities of \exsitu tools.

\subsection{Collaborative Approaches to AR Experience Authoring}
Given the complex and multidisciplinary nature of AR experience development, collaboration is integral to the authoring process, as noted by \citet{kraussCurrentPracticesChallenges2021}. Specifically, AR application development typically involves multiple roles, such as interaction designers, content creators, and developers, each contributing distinct expertise, as well as non-technical collaborators like clients or end-users, who can help shape the design and content~\cite{kraussElementsXRPrototyping2022}. Inspired by role-based collaboration, \citet{nebelingXRDirectorRoleBasedCollaborative2020} introduced \textit{XRDirector}, a system that enables users to manipulate virtual objects from different subjective viewpoints in AR and VR. Designed for indoor immersive storytelling, \textit{XRDirector} enabled multiple contributors to author experiences from perspectives aligned with their respective roles.

Beyond role-based collaboration, prior work has also examined different modes of collaboration when authoring AR experiences. \citet{guoBlocksCollaborativePersistent2019} explored synchronous and asynchronous collaboration in AR authoring through their system, \textit{Blocks}. Their findings indicated that synchronous collaboration fostered higher engagement, particularly when users collaboratively created shared AR structures in real time. Conversely, asynchronous collaboration, while offering greater flexibility, introduced challenges related to maintaining awareness of others' contributions.

While prior systems demonstrate different approaches to collaborative authoring, they do not adequately address the challenges of site-specific AR development, particularly in outdoor environments where \exsitu authors must consider \insitu perspectives. In designing \SystemName, we sought to support this workflow by facilitating collaboration between \exsitu developers and \insitu collaborators without requiring specialized technical skills beyond familiarity with handheld AR. Our system leverages subjective views, similar to \textit{XRDirector}, but applies them to gathering \insitu context and feedback in outdoor, site-specific authoring. Additionally, we compare synchronous and asynchronous workflows, as examined in \textit{Blocks}, including specific hypotheses on engagement and task load (\cref{sec:user-study}).

\subsection{Environmental Context Awareness in Collaborative Mixed Reality}
Effective integration of real-world context is crucial for collaborative AR authoring, particularly when bridging \exsitu and \insitu roles. 
While previous work has explored a wide range of methods to represent remote environments, most focused on indoor settings or static outdoor spaces, leaving the dynamic nature of outdoor environments largely unaddressed.

Early systems such as \textit{MARS}~\cite{hollererExploringMARSDeveloping1999} and \textit{Tinmith-Metro}~\cite{piekarskiTinmithMetroNewOutdoor2001} enabled interaction with virtual models of physical spaces in both indoor and outdoor contexts. However, they lacked visual fidelity, real-time updates, and dynamic environmental data, which limited their effectiveness for site-specific AR authoring.

To enhance environmental awareness, more recent approaches utilize higher-fidelity representations such as $360\degree$-video for immersive panoramic views~\cite{kasaharaJackInHeadImmersive2015,speicher360AnywhereMobileAdhoc2018,teo360DropsMixedReality2019}. For example, \textit{360Anywhere}~\cite{speicher360AnywhereMobileAdhoc2018} supports multi-user collaboration with gaze awareness and annotation tools. While offering comprehensive visual coverage, these methods often lack depth information and do not fully capture the spatial complexity of outdoor environments.

Volumetric representations, including static photogrammetry-based 3D reconstructions and fused RGB-D streams, offer richer spatial representations. Photogrammetry-based methods generate detailed 3D models from images~\cite{gaoRealtimeVisualRepresentations2018,numanExploringUserBehaviour2022, tianUsingVirtualReplicas2023}, though their creation is time-consuming and not suitable for capturing dynamic changes. Systems employing fused RGB-D data, such as \textit{Remixed Reality}~\cite{lindlbauerRemixedRealityManipulating2018} and \textit{TransceiVR}~\cite{thoravikumaravelTransceiVRBridgingAsymmetrical2020}, integrate real-time color and depth information for interactive and dynamic 3D scene representations. However, these approaches commonly rely on specific sensors, system configurations, and controlled settings, which can limit their application in outdoor environments.

Combining multiple spatial representation techniques has proven effective in enhancing environmental awareness in collaborative settings~\cite{teo360DropsMixedReality2019,thoravikumaravelLokiFacilitatingRemote2019,youngMobileportationNomadicTelepresence2020,sakashitaSharedNeRFLeveragingPhotorealistic2024,tianUsingVirtualReplicas2023,reynoldsPopMetaverseMultiUserMultiTasking2024,luRevivingEustonArch2023}. For example, \citet{thoravikumaravelLokiFacilitatingRemote2019} introduced \textit{Loki}, integrating live video feeds, 3D models, and spatial annotations for remote instruction in indoor controlled environments. Focused on representing outdoor environments, \citet{duGeolleryMixedReality2019} projected \textit{Google Street View} imagery onto building geometries derived from \textit{OpenStreetMap} to place and view geotagged information in \textit{Geollery}. While effective within their respective contexts, these systems generally do not account for the challenges posed by highly dynamic, large-scale outdoor environments.

Building on prior work, we aim to address these challenges by enhancing environmental awareness in outdoor environments through \SystemName. We integrate multiple forms of spatial capture methods to supplement pre-captured 3D meshes of outdoor locations used for authoring site-specific AR content. Specifically, \SystemName enables \insitu users to capture single-frame RGB-D data for detailed short-range capture and coarse 3D meshes for broader geometric context. This approach provides \exsitu authors with up-to-date, targeted, real-world spatial context to incorporate into their authoring process.

While several individual components of our proposed system---such as RGB-D capture, mesh reconstruction, and annotation tools---have been explored in prior research, their combined application in highly dynamic, large, and diverse outdoor environments within an AR authoring context represents a novel exploration. By integrating these spatial capture methods into a collaborative workflow, \SystemName aims to bridge the gap between \exsitu and \insitu roles, enhancing environmental awareness and supporting effective real-time collaboration in developing site-specific AR experiences.

\section{Formative Study}\label{sec:formative-study}
To inform the design of \SystemName, we conducted formative surveys and interviews with experienced site-specific AR developers, artists, designers, and testers. Inspired by other investigations into AR development workflows~\cite{speicherXDARChallengesOpportunities2018,ashtariCreatingAugmentedVirtual2020,kraussCurrentPracticesChallenges2021}, we sought to determine current workflows and key challenges of outdoor site-specific AR experience development.

\subsection{Method \& Participants}\label{sec:formative-method-participants}
We recruited 25 survey respondents through internal mailing lists, AR development social media channels, and targeted emails to professional outdoor AR agencies. The survey included both quantitative and qualitative questions about the projects respondents had worked on, the tools they used, and their typical workflow. One survey question prompted respondents to list issues encountered during \insitu testing. To provide an initial set of response options, we defined a set of issue types through a brainstorming session and included an open-ended question for respondents to report additional issues they had come across.

The resulting issue types were as follows:
\begin{enumerate*}[label=(\Alph*)]
    \item Physical constraints (\eg, blocked paths and restricted areas),
    \item User safety (\eg, hazardous environments),
    \item Misaligned AR elements (\eg, misplaced anchors, occlusion, or perspective issues),
    \item User context (\eg, environmental noise or lighting conditions),
    \item Hardware performance (\eg, device processing and rendering power limitations),
    \item Registration issues (\eg, localization and tracking errors),
    \item User flow (\eg, user position or perspective affecting AR interaction),
    \item Socio-cultural appropriateness of the area or experience, and
    \item Semantic interaction (\eg, whether the system correctly recognizes objects in the world).
\end{enumerate*}
We refer to \cref{fig:issue-types} for visual illustrations of these issues.

Five of the survey respondents agreed to complete an interview, in which we asked semi-structured follow-up questions to obtain further details. Two authors of this paper conducted a reflexive thematic analysis to analyze the qualitative results. After developing initial codes, the authors iteratively refined the codebook while reviewing transcripts, following the methods described by \citet{braunThematicAnalysis2019}. Each interview participant received a gift card valued at £25, and survey respondents had a 1-in-10 chance of winning a gift card valued at £30. The survey and interview materials are available in the Supplementary Materials.

\subsection{Findings}
We structure our findings by first outlining typical workflows, followed by a review of major hurdles, and concluding with a discussion of future tools that participants considered useful. These findings inform our understanding of how users typically develop site-specific AR, support us to establish requirements for our proposed system (\cref{sec:system}) and a baseline for our study (\cref{sec:user-study}).

\subsubsection{Workflow walkthrough and tools}
One of the first steps participants took was to \textbf{acquire context about the target location} for their AR experience. This typically involved downloading publicly available pre-captured meshes or visiting the site to scan the location themselves. Participants primarily used \textit{Niantic Geospatial Browser}~\cite{nianticGeospatialBrowser} (64.0\%) or \textit{Google Geospatial Creator}~\cite{googleGeospatialCreator} (28.0\%) to obtain pre-captured meshes, which provided spatial information about the location to support the design and development process. We refer to these pre-captured meshes as \locMeshes throughout the rest of this paper. These representations helped them place and carefully align virtual content with real-world structures and surfaces. Later, when at the site, end-users could use a VPS to localize and interact with the experience.

Following this, participants typically \textbf{worked \exsitu[ ] to design and build the experience} based on the \locMesh. For development, they most often used \textit{Unity} (96\%) and \textit{Blender} (76\%). Additionally, they employed AR frameworks to support features such as plane detection, occlusion meshes, and semantic segmentation. The most commonly used framework was \textit{Niantic SDK for Unity} (80.0\%), followed by \textit{ARFoundation} (76.0\%). A summary of the tools participants used is available in the Supplementary Materials.

Since \exsitu tools could not provide full contextual awareness of the location (\eg dynamic objects, crowding, and lighting conditions), \textbf{participants or their team members often had to travel back and forth} between their \exsitu development location and the target environment. When \insitu[ ], participants logged information about issues with their AR experience to address them later or relay them to their team. On average, they reported spending 36\% (SD=24.7) of their time testing the experience \insitu[ ] and had between 0 and 12 team members that worked with them (M=4.16; SD=2.97). Notably, one interview participant (P4) mentioned considering bringing their ``computer with [their] phone'' to the site to address alignment issues \insitu[ ] but noted that they ``do not think [they] should'' for practical reasons, such as environmental conditions like weather and crowds, which were also cited by other participants (P1, P2, P5).

\begin{figure}
    \centering
    \includegraphics[width=0.94\linewidth]{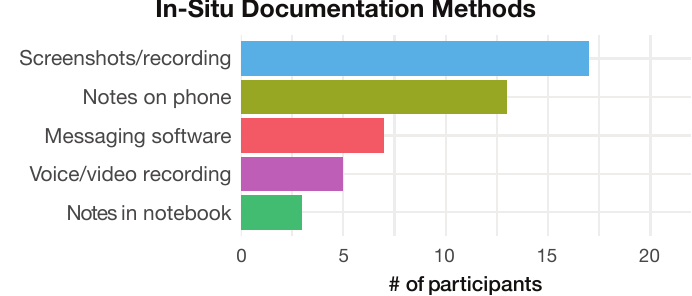}
    \caption{The number of participants who reported using various methods to document issues in their AR experiences when \insitu[ ].}
    \label{fig:formative-documentation}
\end{figure}

\begin{figure}
    \centering
    \includegraphics[width=0.94\linewidth]{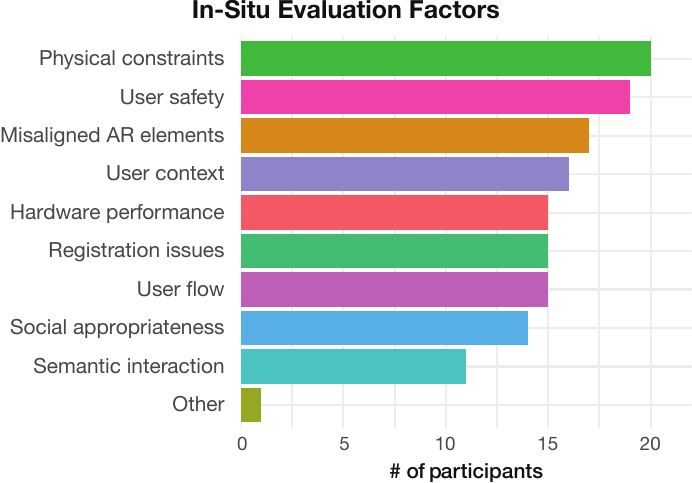}
    \caption{The number of participants who reported looking for various issues in their AR experiences when \insitu[ ]. See \cref{sec:formative-method-participants} for descriptions of the issues.}
    \label{fig:formative-issues}
\end{figure}

In terms of the issues described in \cref{sec:formative-method-participants}, \textit{Physical Constraints}, \textit{User Safety}, and \textit{Misaligned AR Elements} were the most commonly reported\footnote{Only one participant described an additional issue type in the survey, ``Game Flow,'' which we categorize under ``User Flow.''} (see \cref{fig:formative-issues}). To record these issues \insitu[ ], participants typically took screenshots, screen recordings, or typed notes (see \cref{fig:formative-documentation}). Some also received recordings or notes from team members who visited the site. After initial context gathering, participants or their team members continued iterating on their AR experiences, traveling between locations as needed until completion.

\begin{figure*}
    \centering
    \includegraphics[width=\linewidth]{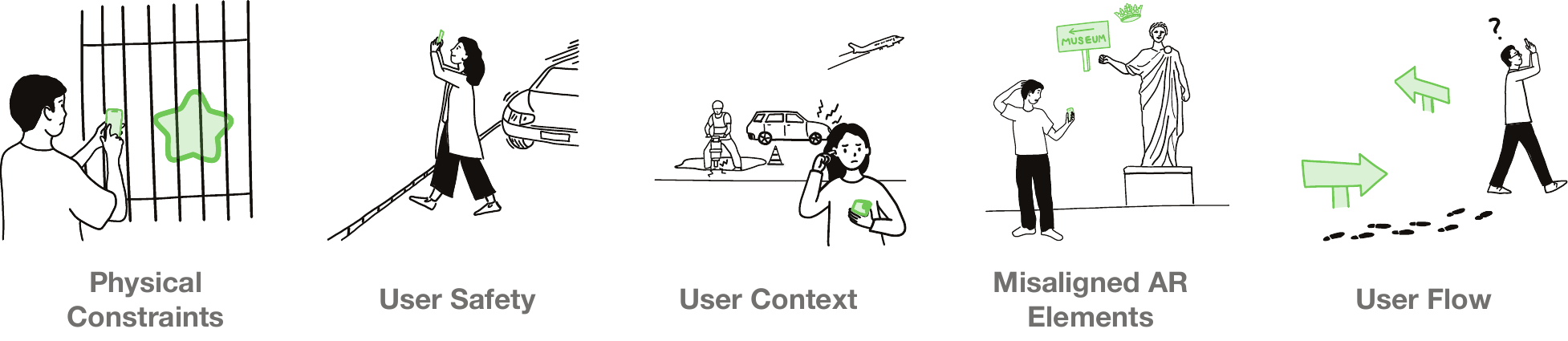}
    \caption{The five main site-specific AR issue types we aimed to address with \SystemName. We refer to \cref{sec:formative-study} for further details on the full range of issues identified in our formative study.}
    \label{fig:issue-types}
\end{figure*}

\subsubsection{Challenges with context capture}
In the interviews, all five participants mentioned \textbf{difficulties with pre-captured meshes}, whether obtained from geospatial browsers or personally scanned. For instance, P2 described pre-captured \locMeshes as ``old'' and not up to date with the current state of the real world. P4 noted that \locMeshes were ``always a challenge because there are mesh holes everywhere [...] so the geometry falls through.'' Similarly, P5 stated that ``the maps are very incomplete,'' but the alternative---scanning the location manually---was ``quite labor intensive.''

Participants also mentioned \textbf{difficulties with scanning}. P1 described how ``often the test scans do not come out great'' and how they sometimes needed to select alternative target locations due to poor scan quality. P5 noted that scan processing times could be lengthy, which reduced iteration efficiency. Two participants also commented on the difficulty of the scanning process due to social awkwardness (P3) or poor weather conditions (P5).

Poor-quality meshes contributed to \textbf{alignment issues} between virtual objects and the real world, both because degraded meshes reduced localization accuracy and because poor mesh geometry hindered precise placement when working \exsitu[ ]. For example, P5 described this limitation: ``It's like you try to figure out where [the virtual object] has to be [ex-situ], and then every time you go [in-situ], your viewpoint is a bit different because you are looking at it from street level or it's slightly tilted [...] we always ended up having to reposition the thing compared to the initial [placement].'' All participants mentioned alignment issues as a recurring problem.

\subsubsection{Missing contextual information}\label{sec:formative-study:missing-contextual-info}
All five participants also mentioned \textbf{challenges in acquiring contextual information missing from meshes}. For example, P3 and P4 described how locations found online might not actually be accessible, or how previously accessible locations might become restricted over time. Three participants mentioned how some locations were hard to conceptualize \exsitu[ ] without visiting in person---either because they were too complex or large (P1) or too cramped (P2, P3).

P1, P2, and P4 mentioned challenges related to \textbf{lighting conditions}, particularly at different times of the day. This concern arose both from an end-user experience perspective (\eg P1: ``You do not want someone staring into the sun'') and in terms of VPS localization accuracy (\eg P4: ``we generally test throughout the day to try to make sure that it is going to work under any lighting''). P1 also highlighted ``audio contamination'' as a factor that could affect the experience, but which could not be discovered without being at the target location. Additionally, P1, P2, and P5 mentioned adverse weather conditions affecting \insitu testing.

One crucial piece of contextual information that all five participants mentioned gathering \insitu[ ] was \textbf{the area's safety}. This included assessing foot and vehicle traffic, as well as the potential for theft. For example, P3 described how even though an area might be plenty of areas where things are walkable [... this still] means very heavy foot traffic, [so] you still have to be careful.'' P1 pointed out that if users are ``getting jostled, or [using] a very expensive phone, then someone could just snatch it.''

All participants also mentioned the need to assess \textbf{the level of crowding at a location}, as high foot traffic could impede both the user experience and VPS localization. P2 even described having to change the target site due to crowding. \textbf{Moving objects} also posed a challenge for both the experience and localization. As objects at the location changed---such as cars, signs, or even ``posters'' (P1)---VPSs struggled to match the environment with the \locMeshes.

\subsubsection{User experience challenges}
All five participants reported \textbf{difficulties identifying end-user experience challenges}, particularly when developing \exsitu[ ]. P1 and P2, for example, emphasized the importance of understanding the ``human scale'' to ensure objects were appropriately reachable and scaled, which was difficult to achieve without being \insitu[ ]. Participants also noted visibility challenges. P1, P2, and P3 described how ``tight FOVs,'' long distances, or sun direction could affect visibility. Gameplay considerations also required \insitu testing; P2 stated that users should not have to run during the experience, while P1 mentioned ensuring that gameplay events were not ``too close'' or ``too far away.''

\subsubsection{Workflow challenges}
Due to these difficulties, four of the five participants described \textbf{traveling to target locations as a significant challenge}. They characterized these trips as ``inefficient'' (P1), ``a lot of effort'' (P3), and ``costly'' (P5). These trips could also be extensive, with participants traveling to different cities (P2, P3) and even different countries (P5) for testing.

Other workflow challenges included collecting and sharing feedback from \insitu testing. P1 described difficulties in physically taking notes while testing the experience, explaining issues over audio, and annotating video recordings. P2 and P5 described communication challenges when working with team members from different backgrounds. P2 also mentioned ``hassles'' related to uploading, downloading, and sharing information while \insitu[ ].

\subsubsection{Desired features for future tools}
As participants described these challenges, they also suggested potential solutions. To address poor context capture, P2 and P4 proposed \textbf{methods to tag areas with semantic or other labels}. Others suggested \textbf{collaborative approaches to context gathering}. For instance, P1, P4, and P5 suggested streamlined methods for sharing and viewing \locMeshes with team members. P4 also proposed a way to combine and edit multiple team members' meshes to enhance their accuracy. Additionally, P2 and P4 expressed interest in a system that allowed team members to collaborate \insitu[ ] and \exsitu[ ] via live streaming of location data and voice communication.

Participants also expressed interest in tools that improve the authoring process. Four of the five participants expressed interest in \textbf{editing objects \insitu[ ]}. P2 and P4 suggested \textbf{procedural placement of objects based on semantic data}. P3, P4, and P5 wanted better methods for simulating the site \exsitu[ ], with P5 stating, ``The more you can do without actually being on the spot, the better.''

\begin{figure*}[ht!]
    \centering
    \includegraphics[width=\linewidth]{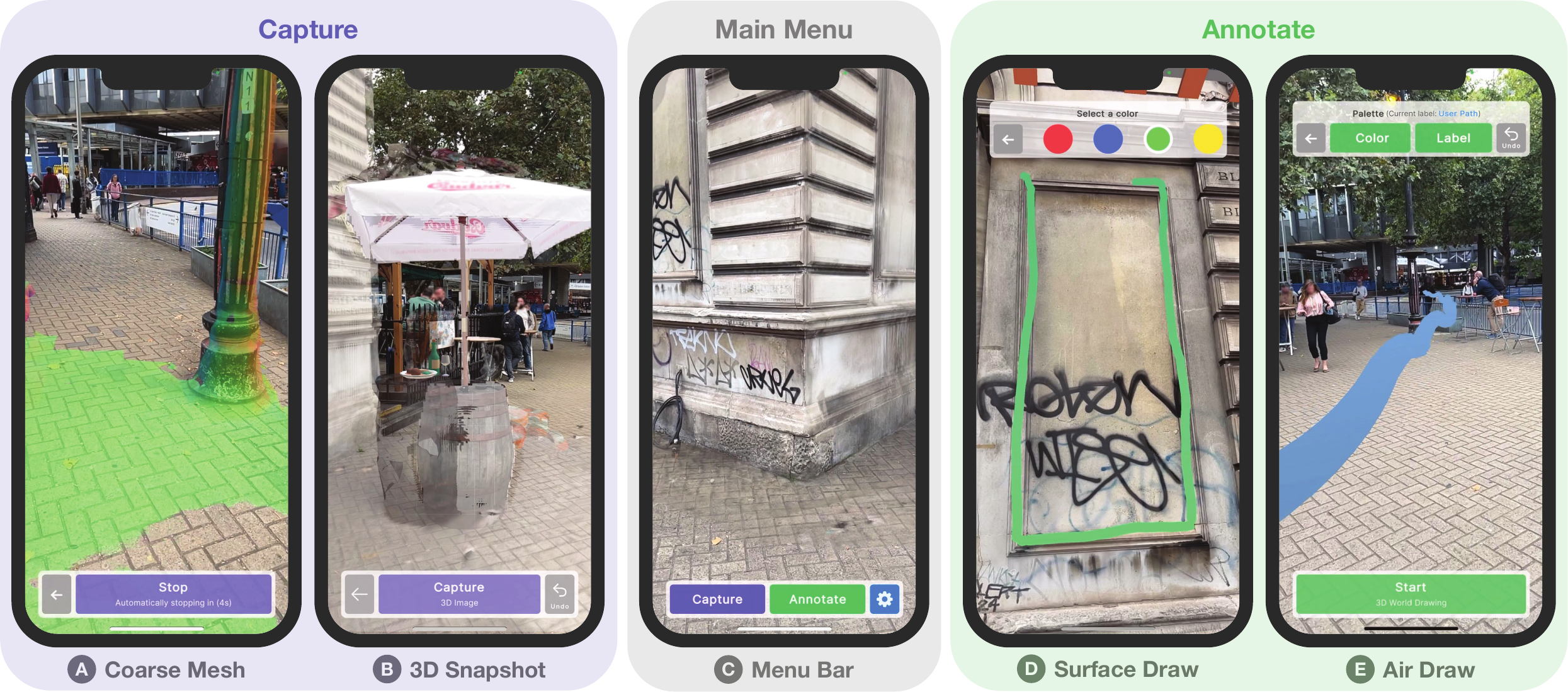}
    \caption{\Insitu user interface of \SystemName. \textsf{(A)} Users can scan the environment to obtain a \textit{Coarse Mesh} of the surroundings; \textsf{(B)} Users can tap \textsf{Capture} to take a \textit{3D Snapshot}; \textsf{(C)} Users can tap \textsf{Capture} or \textsf{Annotate} to access feature sub-menus from the main menu and tap and hold on the screen to spawn a \textit{3D Cursor}; \textsf{(D)} Users can create drawings projected onto surfaces with \textit{Surface Draw}; \textsf{(E)} Users can create a trajectory or a 3D drawing with \textit{Air Draw} by moving their smartphone.}
    \label{fig:insitu-system-overview}
\end{figure*}

\section{\SystemName System}\label{sec:system}
Based on our formative study, we developed \SystemName, a collaborative authoring system for outdoor AR experiences, designed to facilitate real-time interaction between \exsitu developers and \insitu collaborators. 
The system enables \exsitu creators, who typically design and develop the experience within Unity~\cite{unity} asynchronously, to synchronously collaborate with \insitu users who experience the AR content directly in the field. By integrating real-time communication, contextual reference tools, and spatial data capture, \SystemName aims to reduce the need for repeated on-site visits during the iterative design of site-specific AR experiences. While \SystemName supports multiple \exsitu and \insitu users, in this section and our user study, we focus on a two-user scenario in which a single \exsitu user collaborates with one \insitu partner.

\SystemName primarily utilizes two frameworks: the \textit{Niantic SDK for Unity}~\cite{lightship-ardk-niantic}, identified in the formative study as the most widely used outdoor AR development kit, and \textit{Ubiq}~\cite{fristonUbiqSystemBuild2021}, an open-source library for collaborative mixed reality. The \textit{Niantic SDK for Unity} provides VPS localization, depth estimation, meshing, and occlusion handling, whereas \textit{Ubiq} facilitates real-time networking, avatar representation, and peer-to-peer communication, including WebRTC-based audio and video streaming.

\begin{figure*}
    \centering
    \includegraphics[width=\linewidth]{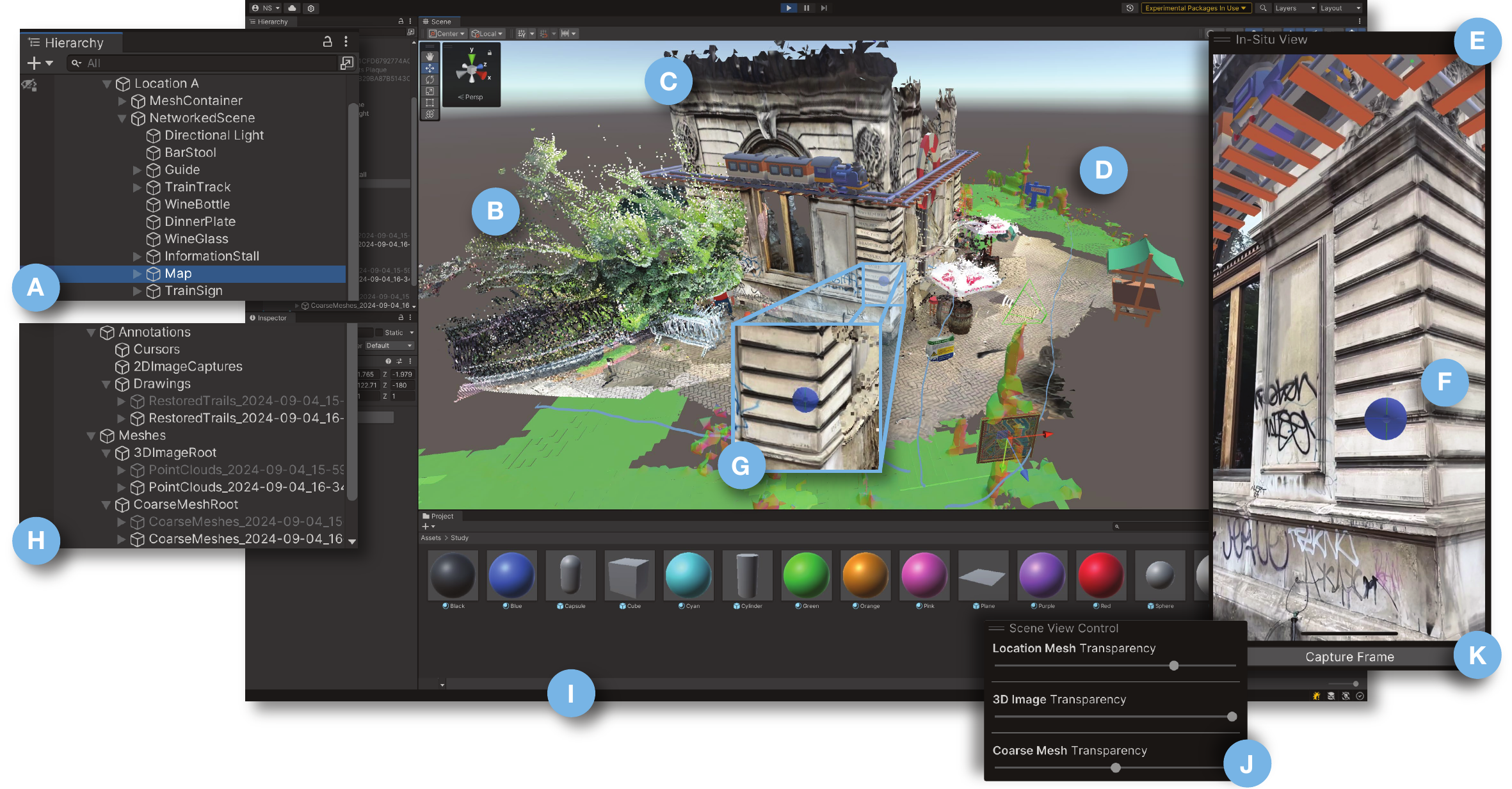}
    \caption{\Exsitu user interface of \SystemName. \textsf{(A)} All objects under \texttt{NetworkedScene} are automatically synchronized between \exsitu and \insitu users; \textsf{(B)} \textit{3D Snapshots} captured by the \insitu user; \textsf{(C)} The \locMesh of \locA; \textsf{(D)} \textit{Coarse Mesh} captured by the \insitu user; \textsf{(E)} Live feed of the \insitu user's screen, including AR content; \textsf{(F)} The \textit{3D Cursor} of the \exsitu user, projected into world space; \textsf{(G)} Close-up of the \textit{3D Cursor} of the \exsitu user as seen in the scene view; \textsf{(H)} List of annotations and spatial captures, persistently saved in the scene for later review; \textsf{(I)} Sample assets that can be added to the scene at runtime.}
    \label{fig:exsitu-system-overview}
\end{figure*}

\subsection{System Requirements}\label{sec:system:requirements}
Based on challenges reported by participants in our formative study (\cref{sec:formative-study}) and previous work, we identified five requirements for the development of \SystemName.

\paragraph{\normalfont\textbf{R1: Reduce the development and testing iteration loop caused by frequent on-site visits}} Our formative study revealed that participants required frequent on-site visits to gather real-world context and feedback. \SystemName aims to reduce the travel burden by allowing \exsitu users to collaborate with \insitu partners synchronously, enabling adjustments to be made remotely based on immediate feedback and live validation.

\paragraph{\normalfont\textbf{R2: Support incorporation of additional or updated real-world spatial information.}} Interviewees highlighted the limitations of static meshes, which are often outdated or incomplete, leading to alignment issues. \SystemName aims to address this by enabling \insitu users to capture and relay real-time updates about the site, such as changes in the environment.

\paragraph{\normalfont\textbf{R3: Facilitate the capture of real-world context beyond spatial data.}} Participants emphasized the importance of capturing additional real-world context, including lighting conditions, pedestrian flow, and safety concerns.
\SystemName incorporates tools that enable \insitu users to capture this contextual information through annotations, real-time audio, snapshots, and a live video feed.

\paragraph{\normalfont\textbf{R4: Enable live persistent scene adjustments and additions during testing and prototyping.}} Participants described having to return to their workplace to make adjustments after \insitu testing as challenging. To mitigate this, \SystemName supports live modifications during testing sessions, allowing developers to make immediate adjustments based on \insitu feedback.

\paragraph{\normalfont\textbf{R5: Integrate captured data into existing development workflows.}} Participants reported difficulties in incorporating \insitu feedback into the virtual scene representations within their development tools, particularly in Unity. \SystemName addresses this by integrating captured spatial data and annotations directly into Unity and anchored to the \locMesh, enabling developers to incorporate feedback without manual import procedures.

\subsection{Scene View and Localization}
\SystemName enables both users to view the same AR content from their respective familiar perspectives. \Insitu users can interact with the AR experience and a \SystemName UI layer within their real-world environment via a smartphone (\cref{fig:insitu-system-overview}), whereas \exsitu users observe the scene remotely through Unity's scene editor (\cref{fig:exsitu-system-overview}).

At launch, \insitu users complete a step-by-step VPS-based localization process to align the AR experience with their surroundings. The system continuously localizes, prompting the \insitu user to confirm alignment by overlaying a semi-transparent pre-captured \locMesh onto the real world. The \insitu user can adjust its transparency to verify overlap before finalizing localization. Once confirmed, the full scene, including the \insitu user's avatar, loads into the \exsitu user's scene view. The \exsitu user can similarly adjust the \locMesh transparency to better view AR elements, spatial captures, and annotations throughout the collaboration process (\cref{fig:exsitu-system-overview}J). If needed, localization can be restarted via the settings panel to correct for drift that may occur during collaboration.

\subsection{Real-Time In-Situ View Streaming and Audio Communication}

\SystemName supports real-time peer-to-peer video and audio communication to facilitate collaboration between users. A peer-to-peer audio channel enables verbal communication. Additionally, a live feed of the \insitu user's screen is streamed to the \exsitu user, who can view this feed in a floating panel within the Unity scene view (\cref{fig:exsitu-system-overview}E). This feed presents both the real-world camera view and the rendered AR elements, enabling the \exsitu user to understand the interaction between AR content and the physical environment.

\Exsitu users can also capture screenshots of the \insitu user's screen, which are saved and displayed in the Unity scene and anchored to the \locMesh (\cref{fig:exsitu-system-overview}K). These captures align with the \insitu user's perspective and can mark specific moments, errors, or areas of interest for later review.

\subsection{Spatial References with 3D Cursors}

To support spatial referencing during the collaborative process, \SystemName incorporates networked \textit{3D Cursors} that each user can control. These cursors are projected into shared world space based on screen inputs. \Insitu users can tap and hold anywhere on their screen to spawn and control a cursor in the 3D environment, while \exsitu users can hover over the live \insitu view panel within Unity to achieve the same. Cursors appear as semi-transparent spheres with axis indicators, aiding spatial referencing. Green represents \insitu cursors, while blue represents \exsitu cursors (see \cref{fig:exsitu-system-overview}F).

Both users can create persistent cursor markers: \exsitu users by clicking within the Unity editor and \insitu users by double-tapping on their screen. This functionality supports alignment tasks and spatial referencing during collaboration.

\subsection{Auto-Networked Objects and Scene Updates}

A core feature of \SystemName is the automatic synchronization of all scene changes (\cref{fig:exsitu-system-overview}A). Any modifications made by the \exsitu user in Unity---such as adjusting the position, rotation, or properties of AR objects---are immediately reflected in the \insitu user's view. This functionality extends Ubiq's networking capabilities to transmit network messages whenever changes in object transforms, materials, or user-controllable serializable script parameters are detected. \SystemName persistently saves all modifications within the Unity scene file, enabling continued iteration and version control after the collaborative session concludes.

\subsection{Spatial Context Capture with Coarse Meshes and 3D Snapshots}
\Insitu users have two tools for capturing additional spatial context to supplement the \exsitu user's representation of the real world: \textit{3D Snapshots}, which are point clouds based on world-projected RGB-D frames, and \textit{Coarse Meshes}, which are untextured 3D meshes.

The \textit{3D Snapshot} feature utilizes the depth submodule of \textit{Niantic SDK} to capture a depth image alongside the current camera frame from a LiDAR-enabled iPhone model. However, we note that \textit{Niantic SDK} also supports 3D data capture on other devices and can extend beyond the LiDAR sensor's range. Following this, the obtained depth data is combined with RGB information to generate a point cloud, which is transmitted to the \exsitu user's Unity scene. There, it is processed and visualized as a colored world-space point cloud anchored to the \locMesh. For \insitu users, the point cloud appears in white and is semi-transparent to provide feedback without distracting from the overall experience. This feature is particularly useful for capturing smaller objects or areas requiring finer detail, such as user interfaces or situated virtual objects that demand precise positioning. 

For larger-scale spatial data capture, \SystemName provides a coarse 3D mesh capturing feature, extending the \textit{Niantic SDK's} meshing submodule. Unlike \textit{3D Snapshots}, the \textit{Coarse Mesh} feature captures the environment's geometry without color information. This method is well-suited for representing larger structures, such as buildings, or open areas, like streets, where surface texture may be less relevant. Mesh updates are streamed to the \exsitu user's Unity editor in real time, with network message coalescing applied to reduce network load. The meshing process automatically terminates after 15 seconds to prevent excessive network usage but can be restarted at any time. The colored mesh material represents surface normals to enhance shape visualization.

Both \textit{3D Snapshots} and \textit{Coarse Meshes} are persistently saved within the Unity scene for future reference, organized into separate timestamped objects for each collaborative session (\cref{fig:exsitu-system-overview}H).

\subsection{Annotation of the Real World with Surface Draw and Air Draw}

\SystemName also enables \insitu users to provide contextual feedback through annotations. The system supports two types of annotations: \textit{Surface Draw} and \textit{Air Draw}. \textit{Surface Draw} projects the \insitu user's sketches onto real-world surfaces using depth estimates, while \textit{Air Draw} allows users to draw freely in 3D space by moving their smartphone through the environment.

Annotations are color-coded and can be tagged with predefined labels, such as \textit{hazard} or \textit{user flow}, or with custom labels which are automatically assigned distinct colors for easy identification. All annotations are streamed to the \exsitu user's Unity editor in real time, where they appear as world-space objects and are saved for future iterations. Like spatial captures, annotations are persistently stored in the Unity scene, allowing contextual information to be revisited and informing subsequent development steps (\cref{fig:exsitu-system-overview}H).

\section{Exploratory User Study}\label{sec:user-study}
\begin{figure*}
    \centering
    \includegraphics[width=\linewidth]{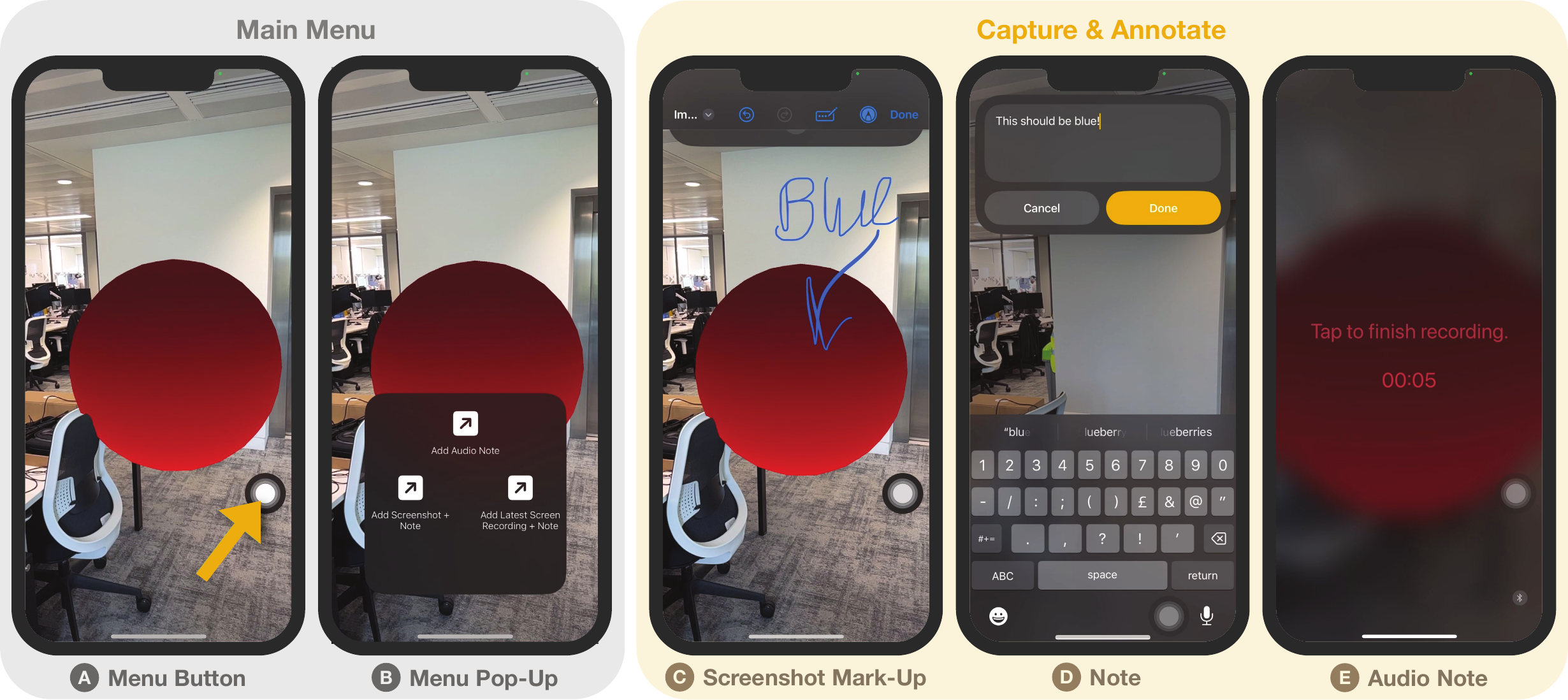}
    \caption{\textit{In-situ} user interface of the baseline system. \textsc{(A)} The menu button (highlighted by a \textcolor[HTML]{F1B302}{yellow} arrow), which is the iOS \textit{AssistiveTouch} button; 
    \textsc{(B)} The menu where users can easily access screenshot, audio note, and screen recording features; \textsc{(C)} Interface for annotating screenshots; \textsc{(D)} Interface for attaching typed notes to screenshots or recordings; \textsc{(E)} Interface for recording audio notes. \textit{Note: the interface for screen recordings is not shown, as the UI is simply a red dot in the upper right corner.}}
    \label{fig:insitu-baseline-system-overview}
\end{figure*}

In this study, we aimed to explore how paired users co-author site-specific AR experiences using \SystemName compared to a typical authoring workflow. In addition to assessing effectiveness and user experience, we sought to identify key directions for future research.

We employed a within-subjects counterbalanced design, comparing \textit{two collaboration modes}: \async (baseline) and \sync (\SystemName, ours). For each session, we assigned one participant to the \exsitu role and the other to the \insitu role, based on their background experience. Participants were compensated with gift cards valued at £40 for their study participation, which lasted around two hours.

To control for order and environmental effects, we counterbalanced both \textit{collaboration mode} (\sync or \async) and \textit{location} (\locA or \locB). Each pair began by completing the task using either the \sync or \async \textit{collaboration mode}, with the \insitu user physically at either \locA or \locB, while the \exsitu user participated remotely. They then completed the task for a second time using the alternate \textit{collaboration mode}, with the \insitu user at the alternate \textit{location}.

We focused on evaluating the below hypotheses, which emerged based on the formative study and prior work~\cite{guoBlocksCollaborativePersistent2019,kraussCurrentPracticesChallenges2021,walkerExperiencingFlowDoing2010,ngSituatedGameLevel2018}:

\begin{hypothesis}\label{h1}
\Exsitu users will report a \textbf{lower} level of task load for \sync{} compared to \async{}.
\end{hypothesis}

\begin{hypothesis}\label{h2}
\Insitu users will report a \textbf{higher} level of task load for \sync{} compared to \async{}.
\end{hypothesis}

\begin{hypothesis}\label{h3}
\Exsitu and \insitu users will report a \textbf{higher} level of engagement for \sync{} compared to \async{}.
\end{hypothesis}

\begin{hypothesis}\label{h4}
\Exsitu users will report a \textbf{higher} level of confidence in authored experiences for \sync{} compared to \async{}.
\end{hypothesis}

\subsection{Baseline}
Our baseline condition, \async, was designed to reflect the typical workflow identified in the formative study (\cref{sec:formative-study}). In this workflow, developers commonly receive or create a set of notes, recordings, and screenshots with insights on improving the AR experience. To align with this, we included these elements in the baseline condition. In our study, the \insitu participant gathered this feedback using a series of custom iPhone \textit{shortcuts}\cite{appleShortcutsUserGuide2024}, activated via the iOS \textit{AssistiveTouch} button. These shortcuts allowed for the creation of voice recordings, annotated screenshots, and screen recordings of the AR application, each automatically saved to the iOS \textit{Notes} app. The user interface of this baseline system is shown in \cref{fig:insitu-baseline-system-overview}.

Once the \insitu participant finished gathering information about the experience, the notes were sent to the \exsitu participant. To recreate a typical developer environment, the baseline system for the \exsitu user was based on Unity~\cite{unity} and \textit{Niantic SDK for Unity}~\cite{lightship-ardk-niantic}. The Unity editor contained the \locMesh, an initial prototype of a site-specific AR experience, and a set of sample assets, including objects and materials from the public domain\footnote{\url{https://kenney.nl/assets/}}.
More information on the study procedure is available in \cref{sec:procedure}.

\subsection{Participants}
We recruited 32 participants (16 pairs) through internal mailing lists and social media platforms. Participants self-reported their gender, with 19 identifying as men and 13 as women. The mean age of participants was 30.16 years (SD = 9.78). Participants were primarily students and researchers from fields related to computer science, virtual and augmented reality, and interactive media, with a few professionals from sectors such as government, consulting, and game development.
For \insitu participants, no prior experience was necessary. For \exsitu participants, basic Unity experience was required. To increase our participant pool and obtain a broad range of experience in Unity---similar to the demographics identified in our formative study---we enabled \exsitu participants to participate remotely from anywhere in the world, using a high-speed internet connection. All participants collaborated in English.

Given the exploratory nature of our study, we selected a sample of 32 participants, consistent with related research on collaborative and authoring systems~\cite{thoravikumaravelTransceiVRBridgingAsymmetrical2020, thoravikumaravelLokiFacilitatingRemote2019, nebelingXRDirectorRoleBasedCollaborative2020}. Moreover, the requirement for participants with Unity experience imposed recruitment constraints, limiting the feasibility of a larger sample~\cite{lakensSampleSizeJustification2022}. Given the study’s scale and exploratory nature, the statistical results should be interpreted with due consideration of potential biases and limitations, as outlined in \cref{sec:discussion:limitations}.

\begin{figure*}
    \centering
    \includegraphics[width=\linewidth]{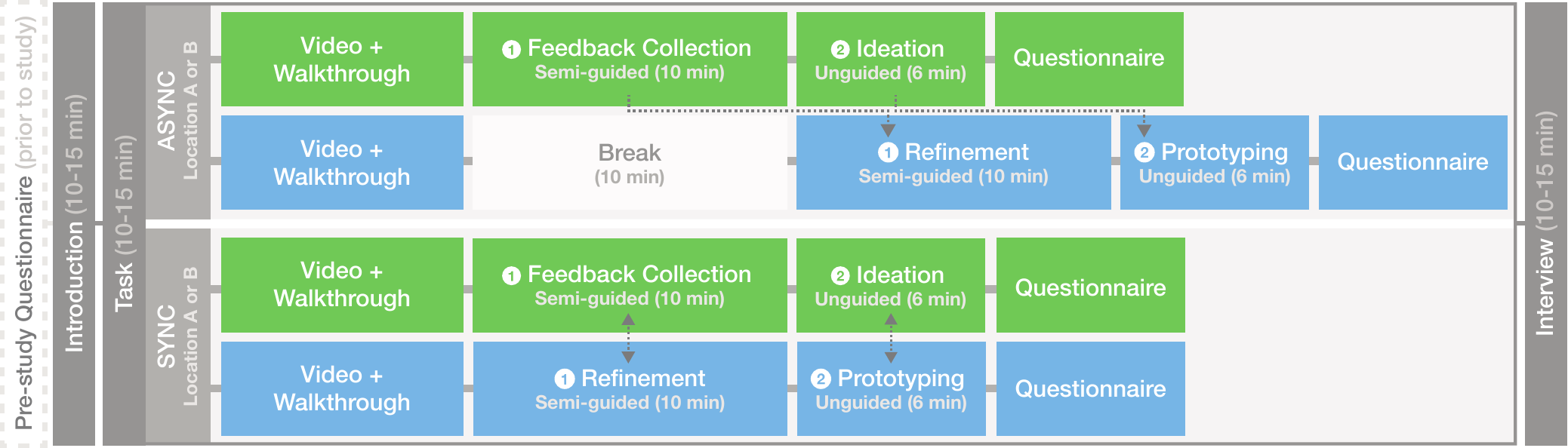}
    \caption{Overview of the study procedure. The process begins with a general introduction (left), followed by the task phase, performed under two conditions (\sync (i.e., \SystemName) and \async), and concludes with a post-study interview (right). \textcolor[HTML]{6DD268}{Green} represents the procedure components of the \insitu participants, \textcolor[HTML]{67B3E6}{blue} represents the procedure components of the \exsitu participants, and \textcolor[HTML]{6e6e6e}{grey} blocks represent components completed by both. \ding{202} represents Phase 1 (Feedback Collection \& Refinement) and \ding{203} represents Phase 2 (Ideation \& Prototyping) of the task. The arrows represent the exchange of task-related information across pairs. Note: both \textit{collaboration modes} (\sync and \async) and \textit{locations} were counterbalanced across pairs.}
    \label{fig:study-procedure}
\end{figure*}

\subsection{Task}
The task involved a two-part collaborative process aimed at improving and expanding predefined site-specific outdoor AR prototype experiences, which are described in \cref{sec:prototype}. 

In \textit{Phase 1: \textbf{Feedback Collection \& Refinement}}, participants focused on identifying and addressing issues within the prototype. The \insitu participant primarily gathered feedback and contextual insights through real-world interaction with the prototype, while the \exsitu participant refined the prototype by implementing necessary corrections based on the feedback provided. In the \async condition, the feedback gathered by the \insitu participant was available to the \exsitu participant \textit{after} the feedback collection phase. To accommodate this order of actions in the \async condition, the \exsitu participant had a short break prior to starting this phase while the \insitu participant gathered feedback. In the \sync condition, this process occurred synchronously.

In \textit{Phase 2: \textbf{Ideation \& Prototyping}}, which lasted six minutes, participants transitioned to brainstorming and implementing extensions or enhancements to the prototype. During the \async condition, the \insitu participant ideated independently, noting ideas and gathering relevant contextual artifacts, which were then shared with the \exsitu participant for prototyping. In contrast, in the \sync condition, ideation and prototyping occurred synchronously.

\subsubsection{Prototype AR experiences}\label{sec:prototype} The prototype experiences used in the study were designed based on the environmental characteristics of the two selected outdoor locations (\locA and \locB), informed by both the \locMesh and supplementary information from \textit{Google Maps} and \textit{Google Street View}. To maintain consistency across locations and ensure generalizability to other site-specific AR applications, the prototypes adhered to a set of predefined design dimensions. These dimensions, grounded in prior work and insights from our formative study, were applied to create similar AR prototypes for each location, ensuring participants could address comparable issues across both locations.

\begin{enumerate}
    \item \textbf{Design Patterns} \textit{used as design guidelines for the representation of each AR element, based on~\cite{leeDesignPatternsSituated2023}:} Glyphs (navigation elements, indicative of actions with spatial alignment to referents), Decals (information related to referent context), Trajectories (\eg, paths, outlines, arrows), Labels (contextual annotations), Ghosts (overlays linked to referents), and Audio (non-visual cues providing additional context).
    
    \item \textbf{Referent Types} \textit{used as design guidelines for the physical referent of each AR element}: Small objects (\eg, mobile elements such as signs or markers), Large objects (\eg, statues, lampposts, or trees), Large planes (\eg, open spaces or flat surfaces), and Building structures (static physical structures with specific entry points or facades).
    
    \item \textbf{Alignment Types} \textit{used as design guidelines for the placement of each AR element}: Overlap (directly aligned with referents), Proximity (placed relative to nearby referents), and Surface (mapped directly onto physical surfaces).
    
    \item \textbf{Issue Types} \textit{used as a design guideline to incorporate common design issues, based on the five most prominent issues identified in \cref{sec:formative-study}:} Physical constraints, User safety, Misaligned AR elements, User context, and User flow, as visualized in \cref{fig:issue-types}.
\end{enumerate}

Both selected locations were within walking distance of University College London (\locA and \locB). \locA, located near a major transit station, featured high foot traffic, a nearby road, and elevated levels of environmental noise. The theme of the AR experience at \locA was ``\textit{Welcoming Tourists to the City},'' reflecting its proximity to a transit hub. \locB was situated along a narrow street with a local cafe, characterized by significant foot and bicycle traffic and lower environmental noise. The theme for this location was ``\textit{Promoting and Celebrating the Upcoming Cookie Party},'' aligning with the atmosphere of nearby restaurants and cafes. For each location, seven issues covering the range of issue types, as previously defined, were identified and verified through on-site testing prior to the user study. An overview of the prototypes is shown in \cref{fig:study-prototypes}, with detailed descriptions available in the Supplementary Materials.

\begin{figure*}
    \centering
    \includegraphics[width=\linewidth]{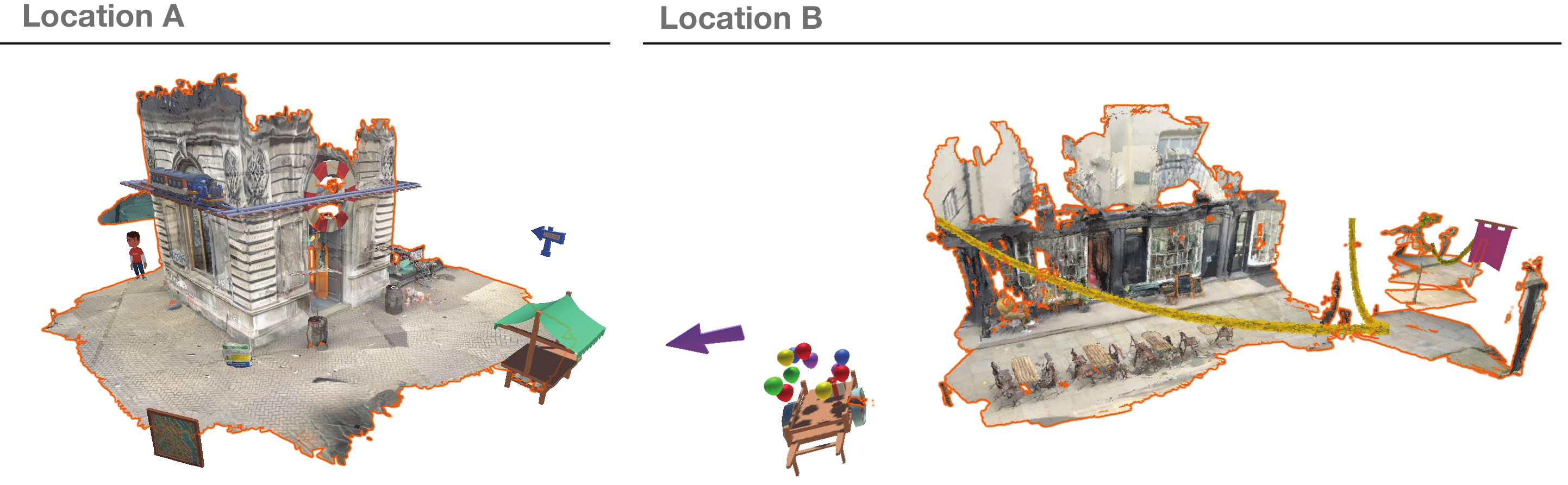}
    \caption{Screenshots of the prototypes designed for \locA and \locB. The \locMesh is outlined in orange, to which the visible virtual low-poly AR elements are anchored.}
    \label{fig:study-prototypes}
\end{figure*}

To offer both participants hints regarding issues in the prototype, both \insitu and \exsitu participants were provided with a list of seven \textit{task clues} in Phase 1. This list contained fictional observations made by ``early testers'' of the prototype experience, offering participants a semi-guided path through the experience while still requiring further investigation to ensure consistency across pairs. Both task clue lists are available in the Supplementary Materials. Participants were allowed to start at any point on the list and could skip or revisit clues as they desired.

\subsection{Procedure}\label{sec:procedure}
An overview of the study procedure is shown in \cref{fig:study-procedure}. Prior to the study session, participants completed a questionnaire covering demographics and their background in smartphone-based AR, 3D game engines, and 3D editors.

For each session, the \insitu participant took part in person using an iPhone 13 Pro, while the \exsitu participant participated remotely by connecting to a PC with all necessary tools pre-installed via the low-latency screen-sharing software \textit{Parsec}\footnote{\url{https://parsec.app}}.

Upon arrival, the \insitu participant was welcomed by an experimenter, while another experimenter connected with the \exsitu participant via video call. The \insitu participant was then brought to a room where they joined the same video call. The experimenter provided a general introduction to the study context and procedure. Before the first condition, the \exsitu participant was guided through a setup document containing log-in information and instructions for forwarding their microphone to the PC. Once the \exsitu participant connected, latency statistics were recorded, ranging between ${\sim}35$–$120\text{ms}$ (including network, decoding, and encoding latency) across participants.

Each pair completed the task twice: once under the \sync and once under the \async \textit{collaboration mode} conditions, with one condition occurring in \locA and the other in \locB. At the start of each condition, participants watched a tutorial video tailored to their role and condition. The \insitu participant then proceeded to either \locA or \locB, assisted by an experimenter, where they received a brief walkthrough of the \sync or \async system. Meanwhile, the \exsitu participant received a similar walkthrough remotely, along with an introduction to the prototype experience, providing background information on its components and initial design decisions.

In the \async condition, during Phase 1 (Feedback Collection \& Refinement), the \insitu participant had ten minutes to gather feedback on the prototype experience and any relevant real-world context necessary to address the issues listed on the task clue list. The \insitu participant then proceeded to Phase 2 (Ideation \& Prototyping), where they collected artifacts related to new ideas for improving or extending the prototype. Concurrently, the \exsitu participant began Phase 1 by refining and correcting the prototype based on the other participant's generated feedback artifacts. After six minutes, the \insitu participant's Phase 2 concluded, and their artifacts were made available to the \exsitu participant for Phase 2, which commenced after their refinement phase (Phase 1) was completed. The \insitu participant then returned indoors, guided by the experimenter, and both participants completed post-task questionnaires separately.

The same process was followed for the \sync condition, except that the \insitu and \exsitu participants progressed through both phases simultaneously. Upon completing both conditions, participants were separately interviewed to discuss their experiences across the conditions.

\subsection{Measures and Analysis}
Each collaborative session was recorded from the perspective of the \exsitu participant. In the \sync condition, this recording included the streamed view of the \insitu participant along with the audio captured from both participants. The recordings were coded to track each participant's feature usage throughout the session. An overview of feature usage is provided in the Supplementary Materials and is discussed in \cref{sec:discussion:environmental_context}.

The post-task questionnaire included the NASA Task Load Index (TLX)~\cite{hartDevelopmentNASATLXTask1988} to assess workload across conditions using a 7-point Likert scale. All TLX items were included for \insitu participants, whereas the item measuring physical demand (``How physically demanding was the task?'') was excluded for \exsitu participants due to their seated, computer-based role in both conditions. An overall task load score was calculated by averaging the responses to the included items.

To measure engagement, we employed the short-form version of the User Engagement Scale (UES-SF)~\cite{obrienPracticalApproachMeasuring2018}, focusing on the subscales of \textit{focused attention}, \textit{perceived usability}, and \textit{reward}. The \textit{aesthetic appeal} subscale was omitted as it was deemed irrelevant. An overall engagement score was obtained by averaging the 5-point Likert scale responses across subscales, adjusting for reverse-coded items.

Additionally, we included several custom items in the post-task questionnaire, which are provided in the Supplementary Materials. For \exsitu participants, custom items assessed their \textit{overall confidence} in the final AR experience as end-users would encounter it \insitu[ ], confidence in the successful execution of task components that influence end-user experience, and specific feedback on the goals of the different task phases. Both participant groups also rated perceived overall task performance using a custom item.

Before concluding the study, both participants were interviewed separately. The post-study interviews focused on evaluating their experiences and preferences regarding the systems and workflows in the \sync and \async conditions. Key topics included overall preference, confidence, communication quality, advantages and limitations of each method, and suggestions for improving system features. The full interview scripts are available in the Supplementary Materials. Two authors of this paper used an affinity diagramming approach to analyze and synthesize themes from the interview transcripts.

\section{Exploratory User Study Results}
Building on the formative study and resulting requirements, we present the results of our evaluation of \SystemName following the user study design detailed in \cref{sec:user-study}. We first report the self-reported questionnaire results, followed by four themes derived from a thematic analysis of interviews with \exsitu and \insitu participants. Plots that show an overview of the usage of specific \SystemName features are provided in the Supplementary Materials.

\subsection{Questionnaire Results}
We conducted statistical analyses using linear mixed-effects models, accounting for participant variability as a random effect and controlling for confounding factors such as task location and prior experience. All models converged, and model assumptions were verified (linearity, normality, and homoscedasticity). Where applicable, we report the estimated coefficients ($\beta$), standard errors (SE), p-values ($p$), and 95\% confidence intervals (CI). Unless stated otherwise, each analysis includes 32 observations ($n$=32). In this section, we provide full details for (marginally) statistically significant results, whereas an overview of all statistical results is available in the Supplementary Materials.

Cronbach's alpha values indicated acceptable internal consistency for both engagement and task load measures. Engagement items yielded alpha values of $0.740$ for \insitu users and $0.764$ for \exsitu users. Task load items had alpha values of $0.812$ and $0.772$ for \insitu and \exsitu users, respectively.

\subsubsection{Engagement}
The analysis revealed a significant effect of \textit{collaboration mode} on self-reported engagement for both \exsitu and \insitu participants. \Exsitu participants reported significantly higher engagement in the \sync condition ($\beta = 0.500$, SE = $0.140$, $p < 0.001$, 95\% CI [$0.226$, $0.774$]). A similar effect was observed for \insitu participants, who reported significantly increased engagement in the \sync condition ($\beta = 0.340$, SE = $0.130$, $p = 0.009$, 95\% CI [$0.085$, $0.595$]). A plot of engagement scores is shown in \cref{fig:engagement-task_load-plot}A.

While no significant effect of \textit{location} was found for \exsitu participants ($p = 0.196$), \insitu participants reported significantly higher engagement at \locB ($\beta = 0.299$, SE = $0.130$, $p = 0.022$, 95\% CI [$0.044$, $0.554$]). \textit{Prior experience} with AR or 3D editing tools did not significantly influence self-reported engagement for either group (\exsitu: $p = 0.093$; \insitu: $p = 0.231$).

\subsubsection{Task load}
\begin{figure}
    \centering
    \includegraphics[width=\linewidth]{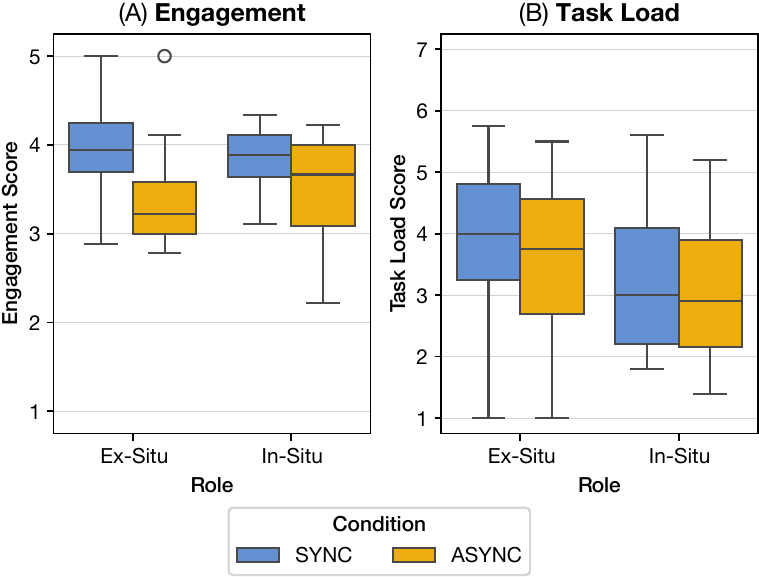}
    \caption{\textsf{(A)} Engagement scores for each role per condition, with significantly higher engagement in the \sync condition for both \insitu and \exsitu participants. \textsf{(B)} Task load scores for each role (\exsitu and \insitu) per condition. No significant differences were found between conditions.}
    \label{fig:engagement-task_load-plot}
\end{figure}

Self-reported task load showed no significant difference between \sync and \async \textit{collaboration modes} for either \exsitu ($\beta = 0.313$, SE = $0.309$, $p = 0.312$) or \insitu participants ($\beta = 0.213$, SE = $0.345$, $p = 0.538$). A plot of task load scores is shown in \cref{fig:engagement-task_load-plot}B.

\textit{Location} did not significantly impact self-reported task load for \exsitu participants ($p = 0.419$). However, \insitu participants reported a marginally significant decrease in task load at \locB compared to \locA ($\beta = -0.637$, SE = $0.345$, $p = 0.065$, 95\% CI [$-1.314$, $0.039$]). \textit{Prior experience} did not significantly affect self-reported task load for \exsitu participants ($p = 0.162$). However, \insitu participants showed a marginally significant association between lower task load and more AR experience ($\beta = -0.320$, SE = $0.176$, $p = 0.070$, 95\% CI [$-0.666$, $0.026$]).

\begin{figure}
    \centering
    \includegraphics[width=\linewidth]{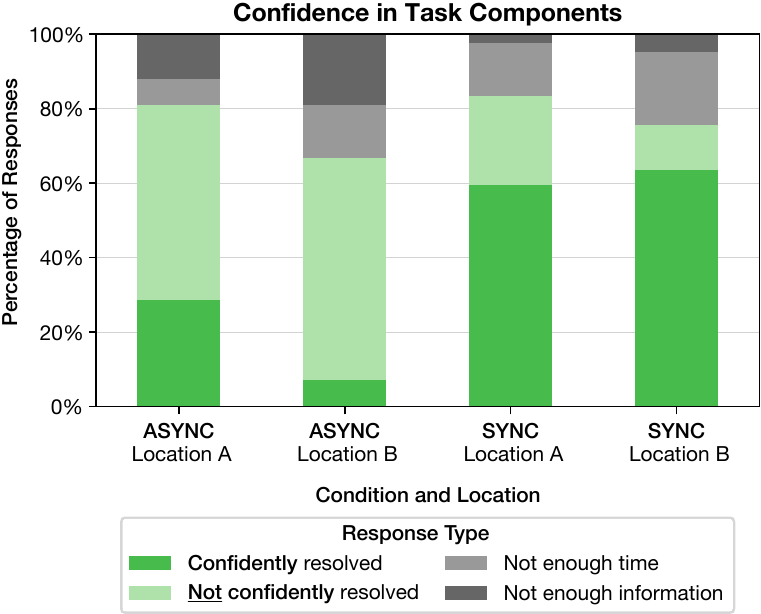}
    \caption{Stacked bar plot illustrating response proportions where participants indicated perceived performance and confidence for each task component across \textit{collaboration mode} and \textit{location} combinations.}
    \label{fig:confidence-task-plot}
\end{figure}

\subsubsection{Confidence in authored results}
Confidence in the number of issues fixed and overall self-reported confidence were both significantly influenced by \textit{collaboration mode}. Participants in the \sync condition reported fixing more issues with confidence than those in the \async condition ($\beta = 3.000$, SE = $0.373$, $p < 0.001$, 95\% CI [$2.270$, $3.730$], $n=24$). Additionally, overall confidence scores were significantly higher in the \sync condition ($\beta = 1.125$, SE = $0.387$, $p = 0.004$, 95\% CI [$0.367$, $1.883$]). An overview of the proportion of participant responses regarding perceived performance and confidence for each task component is shown in \cref{fig:confidence-task-plot}.

A marginal effect of \textit{location} was observed for \exsitu participants, with fewer issues confidently fixed at \locB ($\beta = -0.667$, SE = $0.373$, $p = 0.074$, 95\% CI [$-1.397$, $0.064$], $n=24$). However, \textit{location} did not significantly affect overall confidence scores ($p = 0.747$). \textit{Prior experience} did not significantly impact the number of confidently fixed issues ($p = 0.096$) or overall confidence scores ($p = 0.466$).

\subsection{Interview Results}
At the start of the interview, each participant was asked which system they would prefer to use if they were to perform a similar task again. Among \insitu participants, the majority (12 out of 16) expressed a preference for the \sync condition, citing advantages such as real-time feedback, improved communication, and enhanced engagement. Two participants indicated that their preference would depend on the scenario, while two favored the \async condition, highlighting the ability to focus better without the distractions of real-time interaction.

Similarly, \exsitu participants also largely favored the \sync condition (12 out of 16), with many emphasizing the value of real-time collaboration and immediate feedback. Three participants indicated that their preference depended on the context, and one favored the \async condition due to the slower pace, which allowed more time to refine their contributions.

Our thematic analysis of interview transcripts resulted in four main themes comparing the synchronous \SystemName and asynchronous baseline experiences. We denote \insitu participant quotes with an `I' (\eg, I2) and \exsitu participants with an `E' (\eg, E8).

\subsubsection{\textbf{Collaboration with \insitu users facilitated the integration of real-world context into \exsitu users' design process.}}\label{sec:results:interviews:real-world_context}
In the \async condition, all \exsitu participants struggled to integrate real-world context into their design decisions due to missing information.
\Exsitu users were uncertain about object placement and its interaction with the environment, as they could not verify details without live input. One participant noted, ``I'm not 100\% sure about the real way, if it's low enough or is this blocking other ways because I need to see it from the real end'' (E7). Lacking live visual and spatial information, \exsitu participants relied on limited artifacts provided by \insitu participants, restricting their ability to account for dynamic environmental elements. One participant explained, ``I tried to put the map onto that light post but I'm not sure whether that's on the post or not'' (E7).

In contrast, the \sync condition enabled participants to gain a clearer understanding of real-world context, leading to more informed decision-making in the design process (E12, E10, E9, I5, E17). One \exsitu participant emphasized the importance of real-time observation, particularly for user flow, explaining that although an object appeared well-positioned, real-world feedback revealed that it interfered with pedestrian flow: ``We had a situation where people were walking through one of the objects. It looked like an ideal location for the object on the map [\textit{location mesh}], but in real time and in real life, it wasn't going to work because it was in a walkway'' (E10).

The integration of real-world context through synchronous communication also extended to addressing safety and navigation concerns. One \exsitu participant described uncertainty about whether a virtual cart placed in the scene would interfere with foot and cycling traffic, as there were two real obstacles on either side of it. They noted that this uncertainty was mitigated by being able to check the scene in real time with the \insitu user (E12). Another participant highlighted the value of hearing environmental noise in real time, which influenced decisions about setting audio levels in the AR experience: ``I need to know the population density there, like what's the environment like, should I put the speaker very loud or not at all'' (E9).

Participants also reported that changes in the environment were easier to account for in the \sync condition. One \exsitu user noted that objects had shifted between the time the location mesh was captured and their session, something they would not have noticed without real-time feedback. As they explained, ``The barrel, [...] had actually moved... I wouldn't have been able to tell that from the picture or the scan [location mesh]'' (E17). Following this realization, they asked the \insitu participant to place cursor markers indicating the correct position of the objects on the barrel.

\Exsitu participants also emphasized that synchronous collaboration allowed them to indirectly experience contextual elements of the real-world environment. One participant underscored the importance of experiencing real-world audio and visual elements: ``You can actually hear what they're hearing as well, which is quite important'' (E7). This sense of presence contributed to an increased understanding of spatial relationships within the environment. One participant mentioned, ``I got a much better idea of the space when the other person was there,'' as the \insitu user provided real-time feedback and visual information that could not have been fully captured asynchronously (E16). The combination of real-time video, real-time 3D captures of the environment, and verbal feedback gave \exsitu users a more comprehensive view of the environment, enhancing their understanding of how the design fit within real-world context (E7, E9, E10, E12, E16, E17).

\subsubsection{\textbf{Immediate feedback and changes supported confidence in design decisions, mutual understanding, and perceived accuracy of the outcome.}}\label{sec:results:interviews:immedate_feedback}

All \exsitu participants noted that the \sync condition facilitated more informative feedback and iterative adjustments, with many additionally highlighting that this strengthened their confidence in the authored outcome (I2, E3, E5, E7, I10, E12-17).
The ability to communicate in real time facilitated quicker, informed decisions, with several \exsitu participants noting that the immediate feedback loop significantly reduced the guesswork involved in refining the AR prototype. As E16 stated, ``I think that, very strongly, I'd be more confident that the positions that we put those objects, they actually line up with the real world a lot better.'' This confidence was echoed by other \exsitu participants who mentioned that making the changes in real-time felt ``much, much quicker'' (E3), ``was a lot more dynamic'' (E12), and that ``it felt like, once we were done, it was very complete'' (E10).

The ability to communicate synchronously and see changes immediately not only built confidence but also improved mutual understanding among pairs. Participants emphasized that the visual nature of the real-time interaction removed ambiguity, leading to more accurate placement of virtual objects. Participant I14 noted that discrepancies were minimized: ``I don't think we had a lot of discrepancies in what we discussed about [in the \sync condition], which means probably we had a good sync about the scene.''

\begin{figure*}
    \centering
    \includegraphics[width=\linewidth]{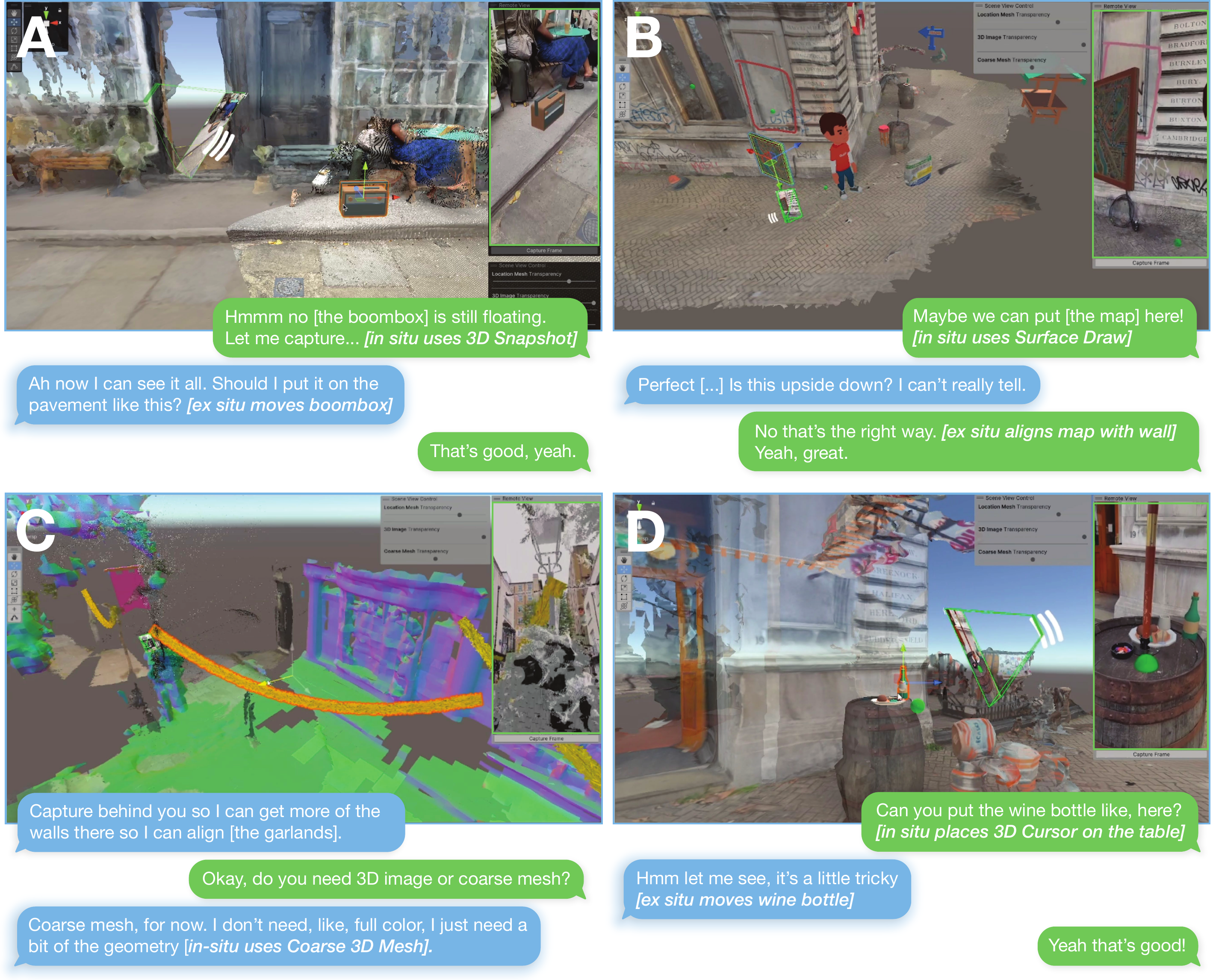}
    \caption{Overview of \SystemName feature usage during Phase 1, shown as \textit{ex-situ} perspective screenshots. Participant conversations are shown in color-coded speech bubbles: \insitu (\textcolor[HTML]{6DD268}{green}) and \exsitu (\textcolor[HTML]{67B3E6}{blue}). Speech bubbles with a \textit{glow} indicate utterances made at the moment of the screenshot. \textsf{(A)} Alignment of a boombox based on spatial context captured using the \textit{3D Snapshot} feature; \textsf{(B)} The \textit{ex-situ} participant moving the map to a position on the wall as specified by the \textit{in-situ} participant through \textit{Surface Drawing}; \textsf{(C)} Alignment of a garland to a previously unmapped region of the street using the \textit{Coarse 3D Mesh} feature; \textsf{(D)} Alignment of misplaced food items based on \textit{in-situ} input using the \textit{3D Cursor}.}
    \label{fig:example-overview}
\end{figure*}

Beyond the real-time feed, \textsc{CoCreatAR}'s additional features --- such as the \textit{3D Cursor}, annotation tools, and the ability to capture \textit{Coarse Meshes} and \textit{3D Snapshots} --- enhanced mutual understanding between \exsitu and \insitu users. \Cref{fig:example-overview} presents examples of feature usage throughout the user study. Participant I16 noted that using the annotation tool to mark exact locations allowed them to ``draw exactly where'' changes were needed, acting as ``the bridge between being there in reality and the scene that [\exsitu user] is seeing.'' Similarly, I5 emphasized that ``drawing in real time is much easier than just explaining it,'' particularly when identifying hazardous areas. The 3D cursor helped reduce ambiguity by allowing participants to visually reference key elements and create placeholders for alignment (I2, I5, I6, I8–10, I12, I14, I16–17, E6, E8, E10–12, E14, E16–17). Additionally, E3 highlighted how the ability to capture \textit{Coarse Meshes} and \textit{3D Snapshots} enabled the \insitu user ``to add more detail to the scene,'' helping \exsitu users make more informed design decisions.

In contrast, the \async condition often resulted in lower confidence due to the lack of immediate feedback and reliance on static information (E9, E12). As E12 described, ``With the asynchronous one, I don't have any information in terms of how much should I move... there's no feedback whatsoever.'' The absence of real-time interaction forced participants to work with limited information, leading to hesitant decisions. E5 emphasized this challenge: ``There was some ineffective information for me in the notes,'' indicating that static instructions without live verification did not provide the necessary context for accurate design decisions. This lack of real-time verification made it difficult to gauge whether changes were correctly applied. E4 highlighted this issue, stating, ``You can't see the change you made, like, in the real world.'' 

\Insitu participants also expressed uncertainty about whether they had captured the right content or communicated their observations clearly (I6, I12, I14, I17). One \insitu participant stated, ``I wasn't sure it was clear enough'' (I14), while another noted that the lack of immediate dialogue required them to provide excessive detail, which still might not ensure accurate interpretation (I17).

\subsubsection{\textbf{Synchronous authoring increased engagement and encouraged creative exploration through collaborative interaction.}}\label{sec:results:interviews:engagement}

In the \sync condition, participants frequently reported higher levels of engagement, often attributing this to the sense of real-time collaboration and mutual decision-making (I3, I8, I12, I15, E3, E5, E12, E14, E17). 
For instance, I10 described the synchronous condition as ``a lot more fun and engaging'' due to the opportunity to work together with another person. Similarly, I14 noted that the synchronous session was ``more enjoyable'' and led to ``higher engagement'' because of the collaborative nature of the task. This sentiment was echoed by E17, highlighting the satisfaction of ``working together'' and described the experience as akin to ``live game testing,'' suggesting that seeing immediate reactions and feedback from the \insitu user created a more interactive and stimulating process. The ability to see their ideas come to life in real time enriched the creative aspect of the experience, as noted by multiple participants (E9, I17).

Synchronous collaboration also drove a sense of teamwork and shared ownership of the final result, contributing to a more engaging authoring process (E7, I10, I12, E16). 
One participant explained that ``it was fun to chat to another person while I was doing it,'' which made the task feel more like a collaborative endeavor rather than an individual effort (I6). This increased engagement was reflected in the enthusiasm with which participants approached the synchronous conditions, with one participant stating that they felt ``buzzing'' with excitement after completing the task together with their collaborator (I12).

In contrast, the \async condition was perceived as less engaging by the majority of participants (I3, I4, E4, I6, E7, I8–I10, I12, E12, I14–I17, E14, E16). 
Participant I12 compared the two conditions, explaining that the asynchronous condition felt ``more like a task as opposed to fun,'' a sentiment echoed by other participants who described the asynchronous process as more isolated and procedural (I4, I7, E14, I17). E12, for example, mentioned that the asynchronous condition felt ``quicker but less satisfying,'' as there was no immediate feedback or dynamic interaction. 

\Insitu participants frequently expressed dissatisfaction with working toward a goal without seeing the final result (I3, I4, I5, I6, I7, I12, I16). 
For example, E16 stated, ``It was a weird feeling leaving and then not seeing it change.'' The lack of real-time collaboration in the \async condition also constrained opportunities for creative exploration. One participant explained that without the ability to brainstorm with another person, the task became ``more about checking a box'' rather than experimenting with new ideas (I10).

Collaborative brainstorming emerged as a key driver of engagement in the synchronous sessions. Participants reported that having someone to exchange ideas with in real time not only enhanced the creative process but also led to more diverse and spontaneous solutions (I9, I10, I17, E14). Participant I9 emphasized the ease of ``drafting concepts'' together in the synchronous condition, stating that ``we could brainstorm easily'' and quickly iterate on suggestions. 
These interactions not only facilitated creative exploration but also enhanced the sense of shared authorship, which several participants valued (I17, E10).

\subsubsection{\textbf{Multitasking overwhelmed some participants in synchronous collaboration, while asynchronous workflows were seen as more suitable for certain scenarios.}}\label{sec:results:interviews:mult-tasking}

Although the synchronous condition generally resulted in increased engagement and collaboration, some participants reported that the increased complexity of multitasking within the synchronous workflow was overwhelming (I8, E16, I17). Some participants also noted that the \sync condition increased pressure, as they felt their collaborators had to wait for them to complete tasks (I4, E12, E14). 

Participant E16 pointed out that managing multiple tasks simultaneously --- such as communicating with the \insitu user, observing the scene, and making design adjustments --- led to cognitive overload: ``I felt like I was doing too many things at once. Talking to [the \insitu user] and trying to watch the environment was sometimes just too much'' (E16). Similarly, I10 expressed that for users unfamiliar with the system, synchronous interactions might feel overwhelming, particularly due to the need to navigate while processing real-time feedback: ``For someone who's not experienced, it was just a lot. You're trying to follow directions, but there's so much going on'' (I10). This aligns with the feedback of several other participants less familiar with AR, who wished they had more time to practice ahead of the task (I3, I6, I10, I13, E10, E11, E16).

Several participants indicated that the synchronous condition was ideal for scenarios where immediate feedback was critical, but the asynchronous condition was more suitable when working in overwhelming environments. Participant I10 described how synchronous interaction could become overwhelming in crowded environments with high noise levels, explaining that ``if you're dealing with a busy place and someone is talking in your ear, it gets really overwhelming'' (I10).

Participants also recognized that both synchronous and asynchronous workflows had their strengths depending on the scenario. While some participants felt overwhelmed by multitasking in the synchronous condition, they still acknowledged its value for tasks requiring rapid decision-making or creative brainstorming. E12 noted, ``It's definitely harder when you have to do everything at once, but I still think the synchronous one is better when you're trying to bounce ideas off someone'' (E12). Participant I16 described how they viewed the \sync condition as most applicable for complex experiences, whereas the \async condition might be more suitable ``if it was, like, a smaller experience with one object.'' E9 also noted scenarios where asynchronous workflows could be beneficial, such as when \insitu participants already have ``enough information to already produce what I want'' (E9).

Moreover, participants reflected on how asynchronous and synchronous workflows or features could be complementary and applied at different stages of the development process (I5, E15, E16). For example, E15 highlighted a hybrid approach: ``I kind of see it as you start with asynchronous, then you do synchronous to refine it, to collect feedback on your experience.'' Other participants saw potential in a hybrid system that flexibly integrates asynchronous and synchronous workflows while incorporating all of \textsc{CoCreatAR}'s capturing and annotation features (I5, E15) and enabling lightweight \insitu editing as in addition (E16).

\section{Discussion}\label{sec:discussion}
In this section, we reflect on the results of the user study, discuss our findings in the context of our formative study and related work, and offer recommendations for future authoring systems of site-specific outdoor AR experiences.

\subsection{Enhancing Environmental Understanding Through Capture and Communication Tools}\label{sec:discussion:environmental_context}
Our findings indicate that synchronous collaboration using \SystemName significantly enhanced the ability of \exsitu participants to integrate real-world context into their design process. In particular, \exsitu participants in the synchronous condition demonstrated a more comprehensive understanding of environmental dynamics, such as user flow and spatial relationships. In this subsection, we contextualize how each feature group contributed to this outcome, drawing on insights from post-study interviews and observations made when coding feature usage counts (see Supplementary Materials for plots).

Participant feedback emphasized that real-time video and audio communication was the most critical feature of \SystemName. It provided \exsitu participants with immediate visual context from the \insitu user’s perspective and supported stable communication. This was especially valuable when \insitu participants felt overwhelmed, enabling them to carry out the task nonetheless. Notably, background noise captured by the \insitu user’s microphone proved useful for decisions regarding audio components (\cref{sec:results:interviews:real-world_context}) --- an issue that, according to P1 of our formative study, could only be resolved when \insitu[ ] (\cref{sec:formative-study:missing-contextual-info}). Additionally, some \exsitu participants mentioned that background audio increased their sense of presence in the \insitu environment.

\textit{Surface Draw} was commonly used due to its task relevance, as objects were typically tied to physical referents (\eg, the map on the wall in \cref{fig:example-overview}B). \textit{Air Draw} was primarily used to indicate user paths and, occasionally, for the creation of 3D wireframe placeholder objects during brainstorming. Labeled annotations were used less often than color annotations, likely because \exsitu participants were not expected to revisit annotations asynchronously at a later stage during the study, unlike in real-world scenarios where annotations might be revisited over time.

\textit{Coarse Mesh} was often used to capture large areas (e.g., buildings in \cref{fig:example-overview}C), while the \textit{3D Snapshot} feature was favored for specific points of interest (e.g., a curb and table set in \cref{fig:example-overview}A). Although the number of \textit{3D Snapshot} captures was higher overall, the longer capture time required for \textit{Coarse Meshes} balanced the frequency of their use. The lack of color in \textit{Coarse Meshes} was not described as a major issue, though one participant creatively captured over ten \textit{3D Snapshots} to obtain a large colored point cloud. \textit{3D Cursors} were mainly used by \insitu participants for object alignment (\cref{fig:example-overview}D) and for supporting deictic references, while \exsitu participants used \textit{3D Cursors} to guide \insitu users' movements or capture actions.

\paragraph{\textbf{Recommendations for future work}}
Enhancing spatial representations and integrating richer metadata layers could expand \textsc{CoCreatAR}'s potential for complex real-world locations. Future work could leverage 3D Gaussian splatting techniques~\cite{kerbl3DGaussianSplatting2023} and their temporal extensions~\cite{wu4DGaussianSplatting2024,mihajlovicSplatFieldsNeuralGaussian2024} to support more intricate design tasks and precise alignment. Automatically capturing metadata such as pedestrian flow~\cite{sindagiGeneratingHighQualityCrowd2017} or hazards~\cite{suRASSARRoomAccessibility2024} during on-site visits could streamline context handling for both synchronous and asynchronous \exsitu authoring.

Future work could also leverage insights from \citet{fussellCoordinationCommunicationEffects2000} to enhance collaboration in \SystemName. For instance, visualizing collaborators' actions, such as object selection or transformation, could improve task awareness. Alternative \insitu interfaces, including head-mounted displays with larger fields of view, might offer a more comprehensive perspective and facilitate natural interactions. Furthermore, hand tracking could enable intuitive gestures, as observed in our study when participants occasionally employed deictic gestures in front of the device camera~\cite{kimWorldPointFingerPointing2023,fussellGesturesVideoStreams2004}.

\subsection{Task Load and Multitasking Challenges}
While our results regarding task load are inconclusive, they offer early insights into the impact of task context, multitasking demands, and individual experiences. A slight trend visible in \cref{fig:engagement-task_load-plot}B suggests a higher task load in the \sync condition for both roles, although no statistically significant differences were observed across conditions. Notably, qualitative feedback revealed that a subset of participants, particularly those with less prior experience, expressed feeling overwhelmed during the \sync{} condition. This feedback provides preliminary insights that challenge \textbf{Hypothesis~\ref{h1}}, suggesting that \exsitu{} participants may experience higher, rather than lower, task load in the \sync{} condition. At the same time, qualitative observations align with \textbf{Hypothesis~\ref{h2}}, as \insitu{} participants reported multitasking challenges indicative of higher task load in the \sync{} condition. Based on the interviews, we conclude that an overall increase in task load during the \sync{} condition could have arisen from the substantial volume of information flow combined with the demands of concurrent coordination and communication. For \insitu users, this challenge appeared compounded by the need to safely navigate the real world---an activity previously recognized as inherently demanding \cite{makhmutovSafetyRisksLocationBased2021}.

Some participants reported that limited familiarity with the system contributed to their sense of being overwhelmed. We hypothesize that these experiences may be influenced by participants' background knowledge and individual personality traits~\cite{johnBigFiveTrait1999}. As we detail in \cref{sec:discussion:limitations}, follow-up research could explore these hypotheses further in larger-scale studies.

Participants' qualitative feedback also highlighted potential \textit{process losses}, defined as productivity reductions that occur when individuals collaborate synchronously rather than working independently~\cite{steinerGroupProcessProductivity1972}. Specifically, some participants reported that the multitasking demands of the \sync{} condition, such as communicating with their collaborator while performing tasks, led to perceived inefficiencies. Related challenges included waiting for collaborators to complete tasks (\eg, annotation by an \insitu{} user or scene editing by an \exsitu{} user) and feeling pressured to avoid blocking their partner's progress (\cref{sec:results:interviews:mult-tasking}). 
\citet{guoBlocksCollaborativePersistent2019} similarly observed potential process losses in synchronous collaboration during an authoring task, specifically noting that skill mismatches within pairs could limit creativity. While we did not observe direct evidence of this particular form of process loss, we theorize that the clearly defined roles and responsibilities assigned to participants, along with the semi-guided nature of the task, may have mitigated these effects.

\paragraph{\textbf{Recommendations for future work}}
Future work could benefit from more extensive onboarding processes to familiarize users with system features and clarify collaborators' roles. Additionally, future iterations of \SystemName could address challenges related to task load by streamlining workflows. For instance, simplified scene editing for \exsitu{} users could be achieved through modular templates~\cite{jurgelionisShapesMarblesPebbles2012} or AI-driven authoring tools~\cite{seeligerContextAdaptiveVisualCues2024,qianScalARAuthoringSemantically2022,evangelistabeloAUITAdaptiveUser2022}. In parallel, \insitu{} users could benefit from AI-assisted annotation methods, such as transcribed audio notes~\cite{langlotzAudioStickiesVisuallyguided2013,kimWinderLinkingSpeech2021}.

To improve orientation and reduce cognitive load for \insitu{} users in complex environments, future systems might integrate visual aids such as mini-maps~\cite{stoakleyVirtualRealityWIM1995} or directional cues~\cite{leeUserPreferenceNavigation2022}, in addition to enhanced visual indicators for collaborator actions as noted in \cref{sec:discussion:environmental_context}. Lastly, for \exsitu{} users, features such as user-perspective scene rendering~\cite{baricevicHandheldARMagic2012} and selective visualization of information layers~\cite{kerstenUsingTaskContext2006,veasExtendedOverviewTechniques2012} could help reduce visual overload.

\subsection{Engagement and Creative Exploration}\label{sec:discussion:engagement}
Both \exsitu and \insitu participants reported higher engagement in the \sync condition, supporting \textbf{Hypothesis~\ref{h3}}. Real-time interaction supported effective teamwork and creative exploration, which the vast majority of participants described as more enjoyable than the \async workflow. This finding aligns with prior research on \textit{social flow}, which found that interdependent team settings amplify engagement compared to solitary workflows~\cite{walkerExperiencingFlowDoing2010}.

Prior work on \insitu AR game level editing \cite{ngSituatedGameLevel2018} found that users were highly engaged in both creating AR content and observing others interact with it. Similarly, in our study, \insitu users were engaged by experiencing and interacting with the AR content as it materialized in their environment, while \exsitu users likened the process to live game testing, emphasizing how immediate \insitu feedback served as a source of inspiration (\cref{sec:results:interviews:engagement}).

The advantages of synchronous collaboration observed in our study are further supported by findings from \citet{guoBlocksCollaborativePersistent2019}, who identified synchronous authoring as a key driver of engagement and creativity compared to asynchronous authoring. Notably, their work additionally revealed a user preference for \textit{co-located} over \textit{remote} synchronous collaborative authoring, which relied on standard video conferencing. This preference underscores the importance of designing remote collaboration tools that replicate the dynamics of co-located experiences more closely. For instance, I3 expressed a desire to see their collaborator, particularly when working together for the first time.

\paragraph{\textbf{Recommendations for future work}}
While \SystemName currently employs avatars to simulate the embodied perspective of co-located collaboration, future research could investigate enhancing interpersonal connection through spatial video-based avatars, as demonstrated in recent work~\cite{vanukuruDualStreamSpatiallySharing2023,qianChatDirectorEnhancingVideo2024}. Another avenue for future research is to explore how creativity can be supported in larger groups, which offer potential for richer creative outcomes~\cite{paulusGroupCreativityInnovation2003} and greater productivity. Although larger groups may face greater coordination overhead, strategies such as role specialization, as employed by XRDirector~\cite{nebelingXRDirectorRoleBasedCollaborative2020}, could streamline the workflows of larger teams.

\subsection{Confidence in End-User Experience Through In-Situ Feedback}
Our findings demonstrate that participants in the \sync condition reported significantly higher confidence in the authored AR experiences than those in the \async condition, supporting \textbf{Hypothesis~\ref{h4}}. Participant feedback indicated that this confidence stemmed from the ability to iteratively verify the designed experience in real time, ensuring alignment with the spatial and contextual dynamics of the real-world environment. In contrast, participants in the \async condition struggled with uncertainty due to reliance on static data and delayed or incomplete feedback, which often led to hesitation in their design decisions.

This positive influence of representative \insitu feedback aligns with prior research in ubiquitous computing that emphasized the critical role of situated evaluation in achieving a reliable understanding of how applications behave in real-world contexts~\cite{rogersWhyItsWorth2007,crabtreeIntroductionSpecialIssue2013}, an approach that our work both embraces and extends.

Notably, much of this research has focused on asynchronous techniques, such as event logging and recordings, to collect \insitu feedback~\cite{rogersWhyItsWorth2007,nogueiraEffectivenessEmbodiedEvaluation2023}. Our findings contribute a new perspective by highlighting the opportunity and value of evaluating and adapting an application simultaneously, a method that our user study revealed to be particularly effective for the iterative refinement of site-specific outdoor AR experiences.

\paragraph{\textbf{Recommendations for future work}}
Future work could explore how author confidence and the perceived quality of AR end-user experiences evolve over multiple revisits of a particular site, especially under dynamic environmental conditions. Additionally, research could focus on developing quantitative metrics that represent the quality of AR experiences, as explored by ARCHIE~\cite{lehmanARCHIEUserFocusedFramework2020}, to provide authors with additional objective insights. While quantitative metrics typically require controlled environments or ground truth sources, metrics derived from user behaviors (e.g., frequency of manual relocalization triggers) could be a practical alternative. Finally, author confidence could be supported by incorporating constraints for contextual adaptation, such as semantic referents and rules~\cite{qianScalARAuthoringSemantically2022,unityMars}, which have been shown to improve confidence when used alongside virtual simulation environments for indoor AR tutorial authoring~\cite{qianScalARAuthoringSemantically2022}.

\subsection{Enabling Hybrid Approaches to AR Experience Authoring}
Despite the advantages of synchronous collaboration, participants identified some scenarios where asynchronous workflows were desired. In contexts with overwhelming environmental conditions or when tasks did not require immediate feedback, asynchronous methods offered more flexibility and reduced pressure. This insight underscores the importance of providing flexible authoring workflows that can adapt to the diverse needs of authoring teams~\cite{kraussCurrentPracticesChallenges2021}, including collaboration across \textit{place} and \textit{time}~\cite{johansenGroupWareComputerSupport1988}.

\paragraph{\textbf{Recommendations for future work}}
Studying the impact of switching between synchronous and asynchronous modes on collaborative processes and authoring outcomes could reveal opportunities for optimization. Furthermore, to support flexible workflows, future work could explore designing hybrid systems that integrate session management and version control for iterative collaborative authoring~\cite{zhangVRGitVersionControl2023,xiaSpacetimeEnablingFluid2018}. Additionally, techniques to visualize synchronous collaboration sessions as artifacts for asynchronous follow-up work could increase and prolong the value of collaborative sessions~\cite{irlittiChallengesAsynchronousCollaboration2016,wangAgainTogetherSocially2020,choRealityReplayDetectingReplaying2023,wangMeetingBridgesDesigning2024}.

\section{Limitations}\label{sec:discussion:limitations}
While our study provides valuable early insights into synchronous and asynchronous collaborative authoring for site-specific outdoor AR experiences, several limitations should be noted.

First, there was variability between the two locations used in the study, which differed in environmental characteristics such as foot traffic, noise levels, and spatial layout. Although we counterbalanced the assignment of locations to conditions, maintained a consistent task across locations, and did not identify a major impact of location in our analyses, these differences may still have influenced participants' experiences and performance.

Second, the sample size of our user study was small, with 32 participants forming 16 pairs. This limited the power of our statistical analysis to detect certain effects. For instance, while qualitative data indicated that some novice participants reported increased task load, this observation may not have been fully reflected in the quantitative data due to the small sample and variability in participants' experience levels. These findings highlight the need for future research with larger sample sizes and specific target groups to examine the role of prior experience and explore strategies for lowering barriers to AR authoring, in line with the recommendations of \citet{ashtariCreatingAugmentedVirtual2020}.

Third, the novelty of the system compared to current methods may have led to participant response bias~\cite{dellYoursBetterParticipant2012}, potentially influencing measures such as engagement. Longitudinal studies with extended and repeated exposure could help isolate these effects.

Finally, our study was limited to a specific type of site-specific AR experience and a particular set of tasks. Consequently, the findings may not generalize to other types and scales of AR application development, such as those involving larger teams or multiple co-located users collaborating on extensive projects. Future research could examine a broader range of AR authoring scenarios that vary in task type and group size, including those with multiple \insitu and \exsitu users engaged in synchronous distributed authoring (e.g., guided tours over larger areas) and environmental exploration.

\section{Conclusion}
This paper examined the challenges of authoring site-specific outdoor AR experiences, which are often constrained by incomplete and outdated world representations and limited access to evolving real-world conditions. Our formative study revealed that developers and designers frequently encounter these limitations, necessitating costly and time-consuming on-site visits to capture environmental details, assess user flow, and ensure contextual relevance. Based on these insights, we identified key requirements for integrating real-world context into remote authoring workflows, leading to the development of \SystemName, an asymmetric collaborative authoring system that facilitates synchronous collaboration between \exsitu (i.e., \textit{off-site}, \textit{remote}) developers and \insitu (i.e., \textit{on-site}) collaborators.

Our exploratory user study demonstrated that this approach mitigates key challenges by enhancing confidence in authored results, stimulating engagement and creativity, and enabling direct iterative refinements informed by up-to-date environmental data. At the same time, our findings highlight multitasking demands as a challenge in synchronous collaboration and emphasize the need for a balanced integration of synchronous and asynchronous workflows. Situating these findings within the broader landscape of AR authoring and remote collaboration, we provided recommendations for future work. We hope this research motivates further exploration of methods for building, testing, and evaluating site-specific AR experiences, contributing to the future of immersive and contextually grounded interactive applications.

\begin{acks}
We express our gratitude to Charlie Houseago, KP Papangelis, Filipe Gaspar, Kelly Cho, George Ash, Adam Hegedus, Stanimir Vichev, Thomas Hall, and Victor Adrian Prisacariu for their feedback and support during the development of \SystemName. We also thank our study participants for their time and insights, as well as Isabel Kraus-Liang for assistance with study logistics. Our user study (\cref{sec:user-study}) was partially supported by the European Union's Horizon 2020 Research and Innovation program as part of project RISE under grant agreement No. 739578.
\end{acks}

\balance
\bibliographystyle{ACM-Reference-Format}
\bibliography{Zotero-References}

\end{document}